\documentclass{ametsocV6}

\usepackage[hidelinks]{hyperref}
\usepackage{mathtools}
\usepackage{amsmath,rotating}
\usepackage{float}
\usepackage{nomencl}

\newcommand{\f}{\frac}
\newcommand{\pd}{\partial}
\newcommand{\ov}{\overline}
\newcommand{\ub}{\underbracket}

\title{Gravity Wave Interactions in the Stratocumulus-Topped Boundary Layer}

\authors{Arun Balakrishna,\aff{a} Hao Fu,\aff{b} Parviz Moin,\aff{a} Morgan O'Neill\aff{c} \correspondingauthor{Arun Balakrishna, arun.balakrishna23@gmail.com}
}

\affiliation{\aff{a}{Center for Turbulence Research, Stanford University, California, USA}\\
\aff{b}{School of the Atmospheric Sciences, Nanjing University, Jiangsu, China}\\
\aff{c}{Department of Physics, University of Toronto, Ontario, Canada}\\
}

\abstract{This work studies the breakup propensity of the stratocumulus-topped boundary layer (STBL) interacting with gravity waves using large-eddy simulation with a uniform vertical grid of $5$~m and horizontal spacing of $30$~m. A radiative-convective equilibrium (RCE) state is constructed to enforce stationarity in the STBL, and the gravity waves are introduced via a vertical momentum forcing mimicking a packet of plane waves. A nondimensionalization involving the inversion height and mean horizontal base wind as length and velocity scales is proposed to provide a framework to analyze the forcing parameter space. The magnitude of the scaled forcing amplitude ($\mathcal{A}$) is critical in understanding various STBL breakup conditions. Classification of breakup was based on the reduction of the liquid water path for each forced STBL case. We found that breakup did not occur for $\mathcal{A}<1$ and observed modest reductions in cloud for $1<\mathcal{A}<2$, but the deck recovered to the stationary state slowly after the single-period forcing ceased. Fixing $\mathcal{A}\sim 2$ showed that forcings with longer duration and wider locality promote breakup. However, when the forcing is a linear combination of waves of two different periods, the percentage of cleared cloud dramatically increases, though recovery of RCE is still observed in some cases. $\mathcal{A}\geq2.5$ marks a critical threshold by which the STBL breaks up entirely and remains patchy. We further explore the connection between these bulk breakup results and the turbulent state by examining energy budgets and the anisotropy induced by the forcing.}

\begin{document}

\maketitle

\pagebreak

%
%
%
%
%


\section{Introduction}\label{sec:introduction}
Stratocumulus clouds are low-altitude clouds that extensively cover Earth's subtropical and midlatitude oceans. Consequently, stratocumuli play an important role in regulating the radiative-exchange budget of the atmosphere by cooling the surface via reflection of sunlight, while minimally impacting terrestrial longwave radiation \citep{Wood2012Sc}. Previous studies have shown that modest increases in low cloud coverage can theoretically balance greenhouse gas induced warming \citep{randall1984,slingo1990sensitivity,goldblatt}. Given the role stratocumulus and other low-altitude clouds play in regulating the climate, it is important to study the factors that lead to their breakup, which can result in significant warming \citep{ARB24,ARB25}. 

Field campaigns \citep[e.g.,][]{slingo1982aircraft,albrechtFIRE,stevens2003dynamics} of the stratocumulus-topped boundary layer (STBL) have formed the basis for understanding stratocumulus physics. One theory of STBL breakup was attributed to cloud-top entrainment instability \citep[CTEI,][]{randallCTEI,Deardorff1980} where turbulence at cloud top promotes runaway deepening of the cloud layer if the change in equivalent potential temperature outweighed latent heating effects across the inversion. However, subsequent studies of observations \citep{KS88,stevensE} and numerical simulations have systematically shown that CTEI cannot be a mechanism for breakup \citep[see section 5.1 in][and references therein]{Mellado2017Entrain}. Large-eddy simulations have also relied on field campaigns as benchmarks for validating numerical calculations \citep[e.g.,][hereafter S05]{Stevens2005LES}. In S05, an ensemble of sixteen LESs with varying numerical discretizations, turbulence closures, grid spacing, etc.\ was compared to the first research flight (RF01) of the Dynamics and Chemistry of Marine Stratocumulus (DYCOMS-II) field campaign, which took place near the California coast. The suite of simulations captured observed thermodynamic and moisture quantities relatively well, but predictions of vertical velocity flux and skewness exhibited a wider spread. We will utilize the numerical framework established by S05 to study stratocumulus as these clouds lay ``on a knife edge between solid or broken regimes (. . .) [and] are also most susceptible to perturbation (. . .)'' \citep{Stevens2005LES}. 

Documented cases of breakup include the existence of open cellular structures within the cloud deck as well as transition of the STBL to a sporadic cumulus state \citep{AlbrechtASTEX,deRoodeSCTASTEX}. The sustenance of these pockets of open cells (POC) has been linked to precipitation  \citep{EPIC,stevensPOC,wood2011aircraft,smalley2022POC} and weaker entrainment rates \citep{bernerPOC,yamaguchi2013size}. Stratocumulus-to-cumulus transition (SCT), however, is mediated by an increase in sea surface temperature (SST) along the transition direction \citep{pincus}, which results in large surface latent heat fluxes decoupling the STBL from the mixed layer \citep{bretherton1997moisture, Yamaguchi2017}. SCT is also sensitive to atmospheric aerosol injection \citep{chunAerosolSc}. 

STBL breakup can also be induced by external perturbations. Previous modeling work in this vein involves varying the background aerosol concentration profile, which can delay breakup timing \citep{goren19Aerosol}. Thermodynamic forcings have also been explored via the modulation of the base state moisture or temperature profiles \citep{bohnert94thesis,wang2010,mechemThermo}
in addition to climatic forcings like doubling carbon dioxide concentrations within the STBL \citep{bretherton2013mechanisms,brethertonLow,Schneider2019Nature}.

A relatively unexplored perturbation mechanism is atmospheric gravity wave (AGW) passage through the STBL. These waves transfer momentum and energy to the global circulation, exhibit a broadband spectrum of wavelengths and periods, and are generated by a variety of atmospheric sources \citep{AGW}. Recent satellite observations during the VOCALS campaign \citep[e.g.][]{Allen2013,yuterSci} revealed trains of horizontally propagating wave packets that led to the formation of POC. LES is an attractive paradigm to explore the wave parameter space as mean convective, turbulence, and moisture effects can be resolved without excessive resolution to capture dissipation and droplet-scale processes \citep{ARB25}. The first LESs of gravity waves impinging on stratocumulus have stemmed from the VOCALS campaign, where the wavelength of these packets was estimated on the order of $100$~km \citep{Jiang2012,Connolly2013}. Because of the long horizontal wavelength relative to the domain size, previous studies have modeled the wave as a subsidence velocity forcing with a prescribed amplitude and period. \cite{Jiang2012} (hereafter JW12) have shown that these gravity waves, whose period is some fraction of the diurnal cycle, augmented precipitation in the cloud leading to cellularization. 

\cite{Connolly2013} (hereafter C13) further estimated wave properties like amplitude and period from VOCALS brightness temperature traces, but again treated the wave as a large-scale vertical velocity. They investigated the effects of amplitude, wave timing, and cloud condensation nuclei (CCN) concentration on reducing the cloud liquid water path (LWP). Most simulations in C13 showed a delayed reduction in cloud condensate hours after the wave was introduced, with sensitivity to CCN concentration and wave timing due to the interplay between the diurnal cycle and decoupling of the boundary layer, which precludes stationarity and masks the breakup dynamics due to the wave alone.

We seek to conduct an additional parametric study of gravity waves incident on stratocumulus. In our study, the forcing waves of interest are in the short-wavelength/period regime, rather than a globally felt subsidence forcing as in JW12 and C13. These short waves have been documented as long-lasting in cloud boundary layers \citep[with periods $\sim \mathcal{O}(10)$~min,][]{jia19}. We systematically control for nonstationary effects like the diurnal cycle by constructing a novel radiative-convective equilibrium (RCE) framework for the STBL. We also employ fine vertical resolution throughout the STBL to capture entrainment scales \citep{brethertonSmoke}, which, to our knowledge, is the first to do so in wave-STBL flows. This study seeks to thoroughly characterize the wave-forced STBL through detailed examination of wave-mean flow interactions, energy budgets, and the turbulent stress state, thereby furthering our understanding of this potential breakup regime \citep{ARB25}.

This paper is structured as follows: \S \ref{sec:methods} details the methods and summarizes the means by which we obtain a steady-state STBL, \S \ref{sec:results} describes the results from the parameter sweep varying the AGW forcing properties. Lastly, we present a summary and future outlook in \S \ref{sec:conclusion}.

\section{Methodology}\label{sec:methods}
\subsection{Model setup}
We perform cloud-permitting LES with Cloud Model 1 \citep[CM1,][]{bryan2002CM1}. The governing equations for mass conservation and momentum in a compressible framework are as follows in tensor notation:  

\begin{equation}\label{eq: Cont}
\f{\pd \rho^*}{\pd t^*} + \f{\pd}{\pd x_j^*}(\rho^* u^*_j) = 0,
\end{equation}

\begin{equation}\label{eq: Mom}
\f{\pd u^*_i}{\pd t^*} + u^*_j\f{\pd u^*_i}{\pd x^*_j} = -c_p\theta^*_\rho \f{\pd \pi'}{\pd x^*_i} - 2\epsilon_{ijk}\Omega_j(u^*_k-u^{*G}_k) + \f{g}{\theta^*_{\rho 0}}\theta'^*_\rho\delta_{i3} + T^*_i + W^*_i + N^*_i. 
\end{equation}

Here $c_p$ is the dry air specific heat; $\Omega$ is Earth's rotation rate; $\rho^*$ is the density of the moist parcel; $u^*_i$ is the velocity in the three ordinate directions; $\theta^*_\rho = \theta^*(1+q_v/\epsilon)/(1+q_v+q_l)$ is the density potential temperature as a function of the potential temperature and specific humidities for vapor and liquid; $\pi = (p^*/p_0)^{R/c_p}$ is the Exner function; $T^*_i$ is the subgrid scale (SGS) stress term; $W^*_i$ is the contribution due to the externally-imposed large-scale subsidence field; and lastly $N^*_i$ is the damping due to any imposed sponge layer. Primes denote deviations from the base state (indicated by a subscript $0$) and asterisked quantities indicate dimensional variables. See the CM1 \href{https://www2.mmm.ucar.edu/people/bryan/cm1/cm1_equations.pdf}{\textcolor{blue}{documentation}} for a detailed description of the prognostic moisture, temperature, and pressure equations and the modified \citet{Deardorff1980} SGS model utilized in this study \citep{ARB25}.  

A few key forcings are necessary to maintain the structure of the simulated STBL. A mean wind of $u_o = 7$~m/s and $v_o = -5.5$~m/s in the boundary layer was chosen based on observation \citep{Stevens2005LES} and enforced as the geostrophic component, $u^{*G}_k$ in the Coriolis term of Eq. \ref{eq: Mom}. To model the impact of circulation on the boundary layer, a large-scale subsidence velocity is prescribed
\begin{equation}\label{eq: subs}
    w^*_{LS} = -Dz^* 
\end{equation} where $D > 0$ is a constant indicative of the bulk divergence. We choose $D = 3.75\times 10^{-6}$~s$^{-1}$ and this velocity is added as a source term:  

\begin{equation}\label{eq: subSource}
    W^*_i = w^*_{LS}\bigg(\f{\pd \phi}{\pd z^*}\delta_{i1} + \f{\pd \phi}{\pd z^*}\delta_{i2}\bigg)
\end{equation} for $\phi = [u^*,v^*,\theta^*,q_v,q_l]$ following S05 \citep{ARB25}. 

\begin{figure}
    \begin{center}
    \includegraphics[width=0.7\textwidth]{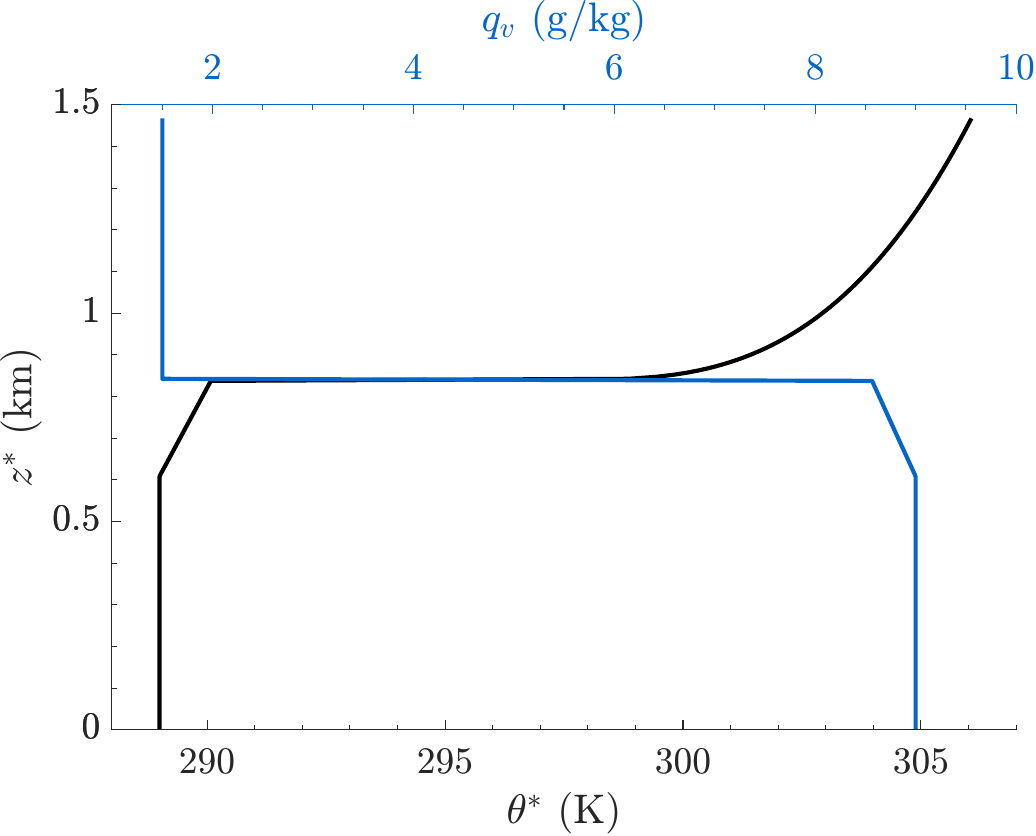}
    \caption{Initial profiles of potential temperature (black) and water vapor specific humidity (blue), which is the same as S05. \label{fig: ScConfig}}
    \end{center}
\end{figure}

We initialize the STBL based on the DYCOMS-II field campaign-derived potential temperature and vapor specific humidity \citep{stevens2003dynamics,Stevens2005LES}. These are plotted in Fig. \ref{fig: ScConfig}; the sharp change in $\theta^*$ at $0.8$~km marks the initial position of the mean inversion height. The domain has a horizontal length of $L_x = L_y = 4\pi$~km and height $L_z = 1.5$~km to support a wide range of wavelengths with potential for horizontal propagation. A grid of $\Delta x = \Delta y = 30.0$~m and $\Delta z = 5$~m leads to approximately 5 grid points across the inversion seen in Fig. \ref{fig: ScConfig}, commensurate with the entrainment resolving guidelines given in \cite{brethertonSmoke}. We use fifth-order finite differencing and CM1's low-storage mixed explicit-implicit RK3-Crank-Nicholson time integration scheme with a maximum time step of $\Delta t=1$~s bounded by a Courant number of $1$ for numerical stability \citep{ARB25}. 

Boundary conditions in the horizontal directions are periodic. The bottom boundary condition is a neutral log law wall model for the ocean surface with a constant sea surface temperature of $292.5$~K, again following S05, along with corresponding surface sensible and latent heat fluxes determined via exchange coefficients for enthalpy and momentum. The top boundary condition is a free-slip condition with a Rayleigh damping layer that relaxes the flow field towards the horizontal average in the upper $300$~m to prevent spurious downward wave reflection \citep{ARB25}. 

\subsection{Gravity wave forcing}\label{sec:GW}
Our gravity wave model is a third imposed forcing acting directly on the vertical momentum equation 
\begin{equation}\label{eq: GWforcing}
    \mathcal{F}_i^* = Ae^{-\big(\f{x^*-x_c}{a}\big)^2-\big(\f{z^*-z_c}{b}\big)^2}\sin(-\omega t^*)\delta_{i3}
\end{equation} where $A$ is the amplitude, $a,b$ are the half-widths in the $x^*$ and $z^*$ directions, $\omega$ is the temporal frequency, and $(x_c,z_c)$ is the center location of the forcing. The form of Eq. \ref{eq: GWforcing} is inspired by a plane wave packet; the Gaussian term also ensures a smooth spatial decay in the forcing. Each parameter constitutes a member of the parameter sweep in this study as $A$ dictates the strength of the forcing, choice of $(x_c,z_c)$ exposes the forcing to differing regions of stratification in the STBL, and $\omega$ implicitly sets the wavenumber spectrum excited by the forcing. In the limit of $a,b \rightarrow \infty$ and oscillatory period of $\sim 1$~hr, the forcing form approaches the (global) subsidence forcing in C13. The use of Eq. \ref{eq: GWforcing} is further validated as this forcing captures the linear dispersion relation of internal gravity waves in a dry, stably stratified setup \citep{ARB25}. See Appendix A for more details. 

\subsection{RCE}\label{sec:RCE} 
A stationary STBL must be established before introducing gravity waves to isolate the forcing effect. The initial state is highly susceptible to strong convective warming, so an adequate radiative-cooling model is needed to reach radiative-convective equilibrium \citep{ARB25}. Based on the domain-averaged turbulent kinetic energy, S05 observed stationarity in their ensemble after 3 hours of model spinup. We propose a stricter measure of stationarity as defined by an unchanging vertical profile of the average equivalent potential temperature 

\begin{equation}\label{eq: theta_eavg}
    \overline{\theta}^*_e(z^*,t^*) = \frac{1}{L_xL_y}\int_{0}^{L_x}\int_{0}^{L_y} \theta^*_e \: dx^* dy^* 
\end{equation} where $\theta_e$ is defined following \cite{emanuel1994atmospheric}. Atmospheric soundings with positive $\pd \theta^*_e/\pd z^*$ indicate an environment stable to moist adiabatic displacement and unstable if this gradient is negative \citep{emanuel1994atmospheric}. Specifically, we require that our RCE convergence criterion $c$ satisfies \begin{equation}\label{eq: RCEcrit}
    c = \max\bigg\{\f{\overline{\theta}^*_e(z^*)|_{t^*+\texttt{T}}-\overline{\theta}^*_e(z^*)|_{t^*}}{\overline{\theta}^*_e(z^*)|_{t^*}}\bigg\} \leq 0.25\%,
\end{equation} where \texttt{T} is some subsequent time after a given averaged profile at time $t^*$. We calculated $c$ using \texttt{T} $\geq 12$~hr. 

\begin{figure}
    \begin{center}
    \includegraphics[width=1\textwidth]{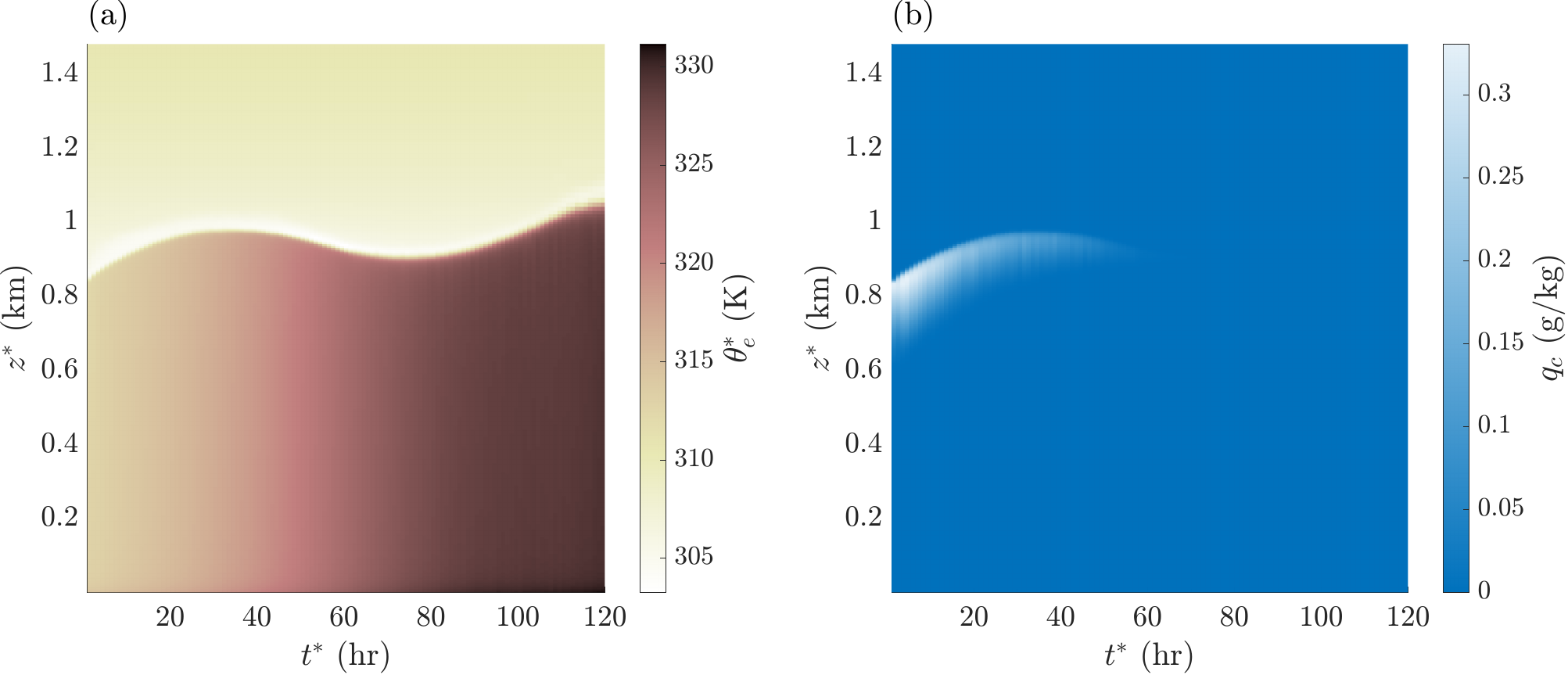}
    \caption{Spatiotemporal contour of (a) equivalent potential temperature and (b) liquid water specific humidity for the \cite{Stevens2005LES} case. \label{fig: S05RCE}}
    \end{center}
\end{figure}

With these definitions in mind, we replicate the STBL calculation performed in S05 with the same spatial discretization as our setup to see if their spinup time is consistent with our RCE/stationarity metric in Eq. \ref{eq: RCEcrit}. The averaged equivalent potential temperature and cloud condensate specific humidity profiles through time for the S05 case are shown in Fig. \ref{fig: S05RCE}. 
 The boundary layer is initially statically unstable, so this promotes convective warming near the surface. A quasi-steady state can be observed over a short time horizon ($t^*<10$~hr), consistent with the reported stationarity time in S05. However as time progresses, an observed thinning of the cloud top is accompanied by a strong increase in $\theta^*_e$ near the surface along with oscillation of the mean inversion layer. Not surprisingly, this leads to no liquid water in the domain after $t^* > 60$~hr \citep{ARB24}. We can better understand this behavior through analysis of S05's radiation model:
\begin{multline}\label{eq: S05radmodel}
    F_{rad} = \ub[1pt]{F_o \exp\bigg[-\kappa \int_{z^*}^{\infty} \rho^* r_l \,dz^*\bigg]}_\text{\clap{cloud-top cooling}} + \ub[1pt]{F_1 \exp\bigg[-\kappa \int_{0}^{z^*} \rho^* r_l \,dz^*\bigg]}_\text{\clap{cloud-base warming}} \\ + \ub[1pt]{\rho_i c_p D \alpha_z\bigg[\f{1}{4}(z^*-z_i)^{4/3}+z_i(z^*-z_i)^{1/3}\bigg]}_\text{\clap{free-troposphere cooling}}. 
\end{multline} The first two liquid water path terms represent cloud-top cooling and cloud-base warming while the third term in Eq. \ref{eq: S05radmodel} is the free-troposphere cooling contribution. As prescribed, the radiative fluxes due to cloud-top cooling and cloud-base warming are exponential functions of $q_l$ leading to markedly smaller radiative fluxes imposed through time and thus are unable to counteract the positive feedback of convective heating aiding evaporation. Dissipation of cloud clearly shows that the S05 STBL was not in RCE and that this radiation model is insufficient to balance the strong convective warming within the boundary layer \citep{ARB23}. 

\begin{figure}
    \begin{center}
    \includegraphics[width=1\textwidth]{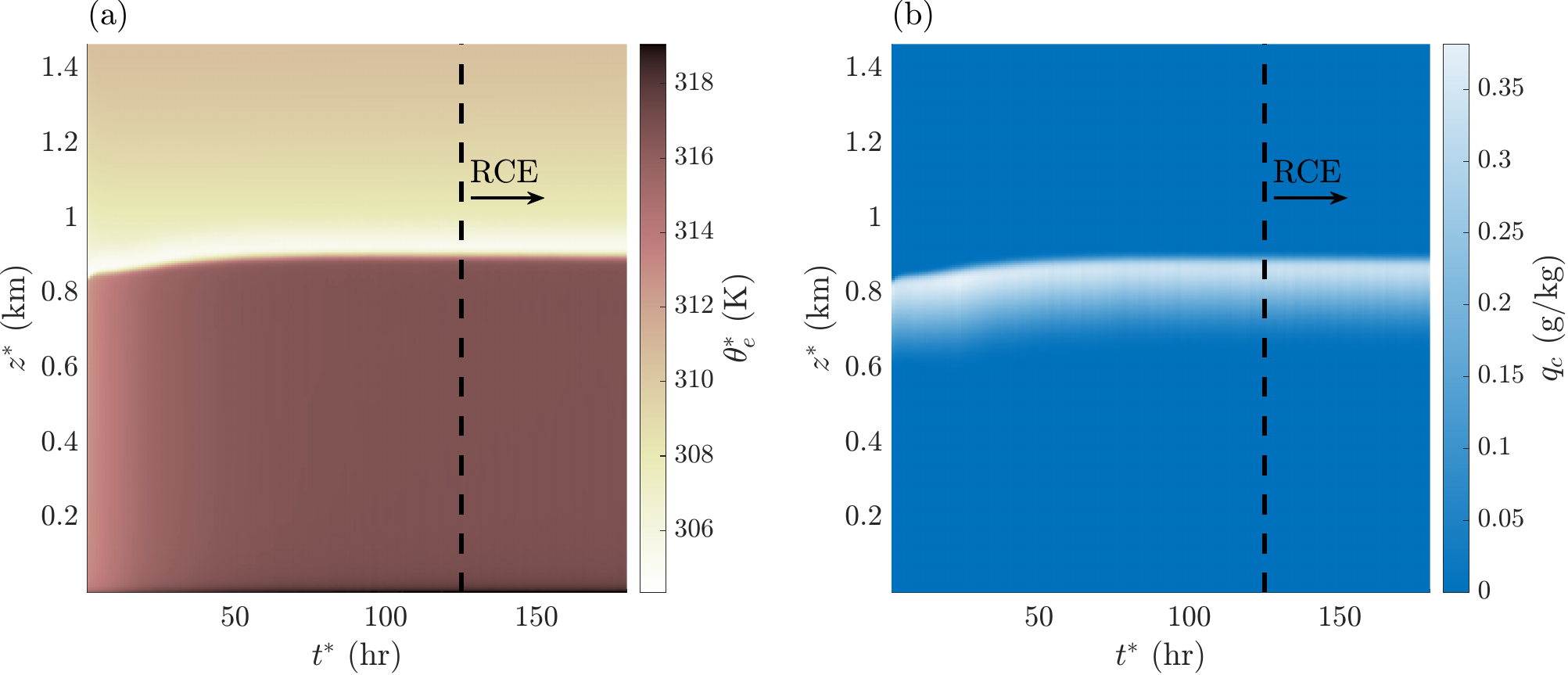}
    \caption{Spatiotemporal contour of (a) equivalent potential temperature and (b) cloud condensate specific humidity for the current setup. The onset of radiative-convective equilibrium is denoted by the black dashed line. \label{fig: RCE}}
    \end{center}
\end{figure}

\begin{figure}
    \begin{center}
    \includegraphics[width=1\textwidth]{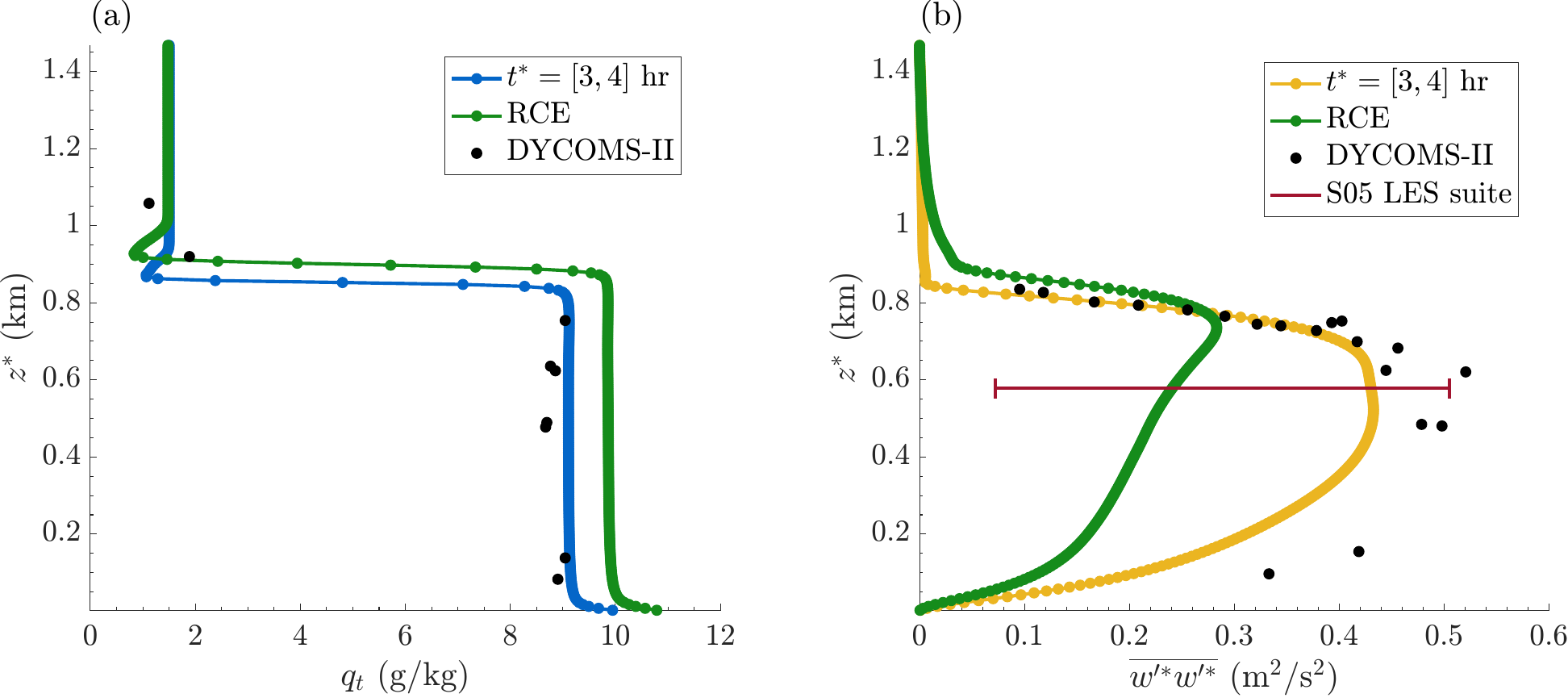}
    \caption{Vertical profiles of a) total water specific humidity and b) vertical velocity flux averaged over the same time window (from 3-4 hr) as S05 which is compared against the profiles averaged during the RCE period and the field data from DYCOMS-II first research flight (digitized from S05). The red ``errorbar'' in b) shows the widest spread in the LES ensemble range from S05 at $578$~m. \label{fig: S05RCEDY}}
    \end{center}
\end{figure}

RCE is crucial as a non-stationary STBL would confound effects of any wave forcing for an idealized mechanism study. Our radiation model retains the LWP radiative terms in Eq. \ref{eq: S05radmodel} but neglects the third term. We instead model free-troposphere cooling with a prescribed newtonian cooling term 
\begin{equation}\label{eq: Qdot}
    \dot{Q}_\theta^* = \f{\theta'^*}{\tau_c}
\end{equation} acting over a constant time scale $\tau_c$. This term is analogous to the clear-sky cooling term in \cite{chung2012steady}, however we constrain Eq. \ref{eq: Qdot} to the region above the inversion layer. Note that radiation models such as RRTMG \citep{IacanoRRTMG} are incompatible with the fine vertical resolution utilized in this work (G. Bryan, private communication). Addition of Eq. \ref{eq: Qdot} to the radiative model does indeed lead to a stationary cloud deck as shown in Fig. \ref{fig: RCE}. Here we have chosen a cooling timescale $\tau_c = 3$~hr. 
 With this free-troposphere cooling model, Fig. \ref{fig: S05RCEDY} shows the total water and vertical velocity flux profiles at the same comparison time as in S05 and once RCE was attained. Both the early total water and momentum flux profiles show fair agreement with the field data, which further supports the choice of $\tau_c$. However, the RCE profile departs from the observations, but this is expected as RCE is an idealized state and not truly representative of the STBL, given the diurnal cycle and variable SSTs. Here we argue that the early transient development of the numerical STBL corresponds with the observations (taken at some finite period in time) on the way to an idealized equilibrium. Note that the attained RCE state still maintains the canonical sharp inversion layer, and the RCE $\ov{w'^*w'^*}$ profile in Fig. \ref{fig: S05RCEDY}b exhibits the same level of fidelity of the S05 LES ensemble as indicated by the red ``error bar''.

\subsection{Recasting the governing equations}
It is helpful to understand the relative magnitude of these forcings in the context of a nondimensionalized momentum equation. We take a velocity scale $U \sim \sqrt{u^{2}_o + v^{2}_o}$, length scale $L \sim \ov{z}_i$ where $\ov{z}_i$ is the average inversion height defined by the $q_t = q_l + q_v = 8$~g/kg isoline \citep{Stevens2005LES}, temperature scale $\theta_s\sim \theta_{\rho0}$, the Coriolis parameter $f$ to normalize $\Omega^*$, and an advective time scale $t_s \sim \ov{z}_i/U$ \citep{ARB25}. Applying the normalization in this manner leads to the dimensionless momentum equations \begin{equation}\label{eq: nondimMom}
    \f{\pd u_i}{\pd t} + u_j\f{\pd u_i}{\pd x_j} = -\mathrm{Ja}_m\theta_\rho \f{\pd \pi'}{\pd x_i} - \f{2}{\mathrm{Ro}}\epsilon_{ijk}\Omega_j (u_k-u^G_k) + \f{1}{\mathrm{Fr}^2}\theta_\rho'\delta_{i3} + \f{\ov{z}_i}{U^2}\bigg[\mathcal{F}^*\delta_{i3} + T^*_i + N^*_i+W^*_i \bigg]
\end{equation} with the three nondimensional parameters being the Froude number ($\mathrm{Fr}$), Rossby number ($\mathrm{Ro}$), and what we term a moist Jakob number\footnote{In multiphase flow, the Jakob number is the ratio of sensible heat to the enthalpy of vaporization} ($\mathrm{Ja}_m$) defined as 
\begin{equation}\label{eq: Ja}
    \mathrm{Ja}_m  = \f{c_p \theta_{\rho 0}}{U^2},
\end{equation}
\begin{equation}\label{eq: Ro}
    \mathrm{Ro}  = \f{U}{f \ov{z}_i}, 
\end{equation} and 
\begin{equation}\label{eq: Fr}
    \mathrm{Fr}  = \f{U}{\sqrt{g\ov{z}_i}}. 
\end{equation}

\begin{table}
    \centering
    \resizebox{0.65\textwidth}{!}{
    \begin{tabular}{ccccc}
         Run ID & $\mathcal{A} = A\ov{z}_i/U^2$ & $z_c/\ov{z}_i$ & $\omega/\omega_{ref} $ & $t_f\omega/2\pi$ \\
         Base &  0.11 & 0.5 &  1 & 1\\
         A1 & 0.56 & 0.5 &  1 & 1\\
         A2 & 1.13 & 0.5 & 1 & 1\\
         A3 & 1.97 & 0.5 & 1 & 1\\
         A3T1 & 1.97 & 0.5 & 1.5 & 1\\
         A3T2 & 1.97 & 0.5 & 3 & 1\\
         A3T3 & 1.97 & 0.5 & 6 & 1\\
         A3CF & 1.97 &  0.83 & 1 & 1\\
         A2B & 1.13 & 0.5 &  (6,1) & (6,1)\\
         A3B & 1.97 & 0.5 &  (6,1) & (6,1)\\
         A4 & 2.82 & 0.5 & 1 & 1\\
    \end{tabular}}
    \caption{Gravity wave forcing parameter sweep runs.}
    \label{tab: sweep}
\end{table}

Table \ref{tab: sweep} summarizes the simulations performed in terms of the nondimensional amplitude, vertical location, frequency, and number of periods forced. The experimental run Base has wave parameters whose values are observationally grounded: we define a reference period of $\mathcal{T}_{ref} = 1.5$~hr (such that $\omega_{ref} = 2\pi/\mathcal{T}_{ref}$) which is the same as the satellite-derived measurements in C13. Choice of Base forcing's amplitude is based on an order-of-magnitude analysis. In C13, the peak forced subsidence velocity occurred at $z_i$ with a corresponding value of $\mathcal{O}(10^{-1})$~m/s. Assuming the acceleration timescale is also $t_s \sim \mathcal{O}(10^1)$~s, this leads to a dimensional amplitude estimate of $10^{-2}$~m/s$^2$ \citep{ARB25}.

\begin{figure}
    \begin{center}
    \includegraphics[width=0.8\textwidth]{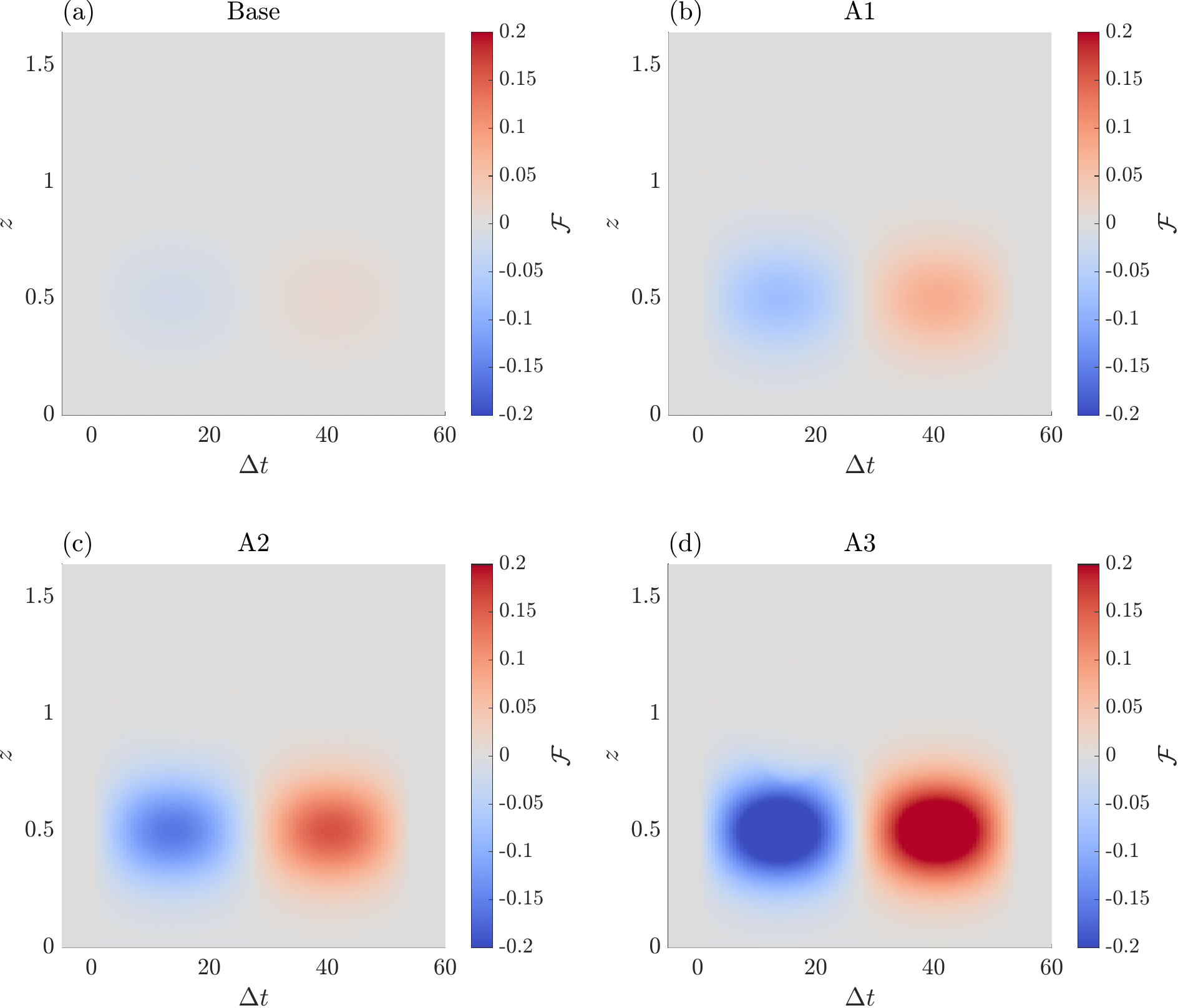}
    \caption{Spatiotemporal contour of the nondimensional forcing for the amplitude simulations: Base, A1, A2, and A3. Note how the A3 forcing is depressed near $\Delta t = 15$ as a result of a lower $z_i$. \label{fig: ampForce}}
    \end{center}
\end{figure}

Subsequent runs involve varying the amplitude (A1-A4), varying the period with fixed amplitude (A3T1-A3T3), constraining the forcing within the cloud region (A3CF), and investigating the effect of a bichromatic forcing (A2B-A3B). The half-widths are chosen to constrain the forcing below the inversion so as not to energize the non-turbulent free-troposphere. The forcing is explicitly set to 0 above $z^*=z_i$, because $z_i$ is dynamic depending on the level of vertical displacement induced by the forcing itself. In the $x$ direction, all experiments utilized $x_c/\ov{z}_i = 3.36$ and $a/\ov{z}_i = 1.12$. This ensures a sufficient distance from the streamwise boundaries allowing the forced waves to propagate horizontally, and keeps the forcing region relatively localized in $x$ over a distance of $2$~km, respectively. In $z$, we center the forcing halfway in the STBL ($z_c/\ov{z}_i = 0.5$) and choose $b/\ov{z}_i = 0.223$ allowing for a smooth decay in $\mathcal{F}$ by $z^*=z_i$. The one exception is for A3CF which had $b/\ov{z}_i = 0.087$ to constrain the forcing to the cloud layer itself. The chosen forcing location represents a theoretical impingement location from a generic wave source like a convective disturbance or orography \citep{sun2015review}. Fig. \ref{fig: ampForce} illustrates a Hovmöller diagram of the dimensionless forcing amplitude parameter space. All runs have an active forcing time of one integer period. Note that for each of the forced calculations, Rayleigh damping on the velocity fields is additionally applied $300$~m from each boundary to prevent re-entrant waves with a relaxation time scale of $30$~s. Note that $\ov{z}_i = 894$~m in the RCE state of the STBL.

\subsection{Base state for wave-forced calculations}
In CM1, the base state is defined via the initial condition. While Fig. \ref{fig: ScConfig} initialized the RCE simulations, we want the RCE profile to be the base state upon which the wave-forcing is introduced and the reference state from which flow deviations are calculated in Eq. \ref{eq: nondimMom}. Hence, a simple restart from the RCE period is not appropriate; rather, the RCE vertical profiles of $\theta$ and $q_v$ are extracted into a custom input sounding profile. In addition, we modify the source code to account for a nonzero base state $q_c$ in order to reestablish the RCE cloud deck. We then run this new calculation now ``anchored'' to the STBL RCE state in Fig. \ref{fig: RCE} for a brief spin-up period because of the introduced transients on account of the horizontal sponge. After the transients in this RCE-anchored STBL are flushed, we impose the various forcings in Table \ref{tab: sweep}. 

\section{Results}\label{sec:results}
\subsection{Gravity wave forcing sweep}
A simple statistic to compare the various cases in the forcing parameter sweep is the percentage of cloud cleared during the simulation time period. A cloud-free column is defined as having an LWP $ < 10^{-3}$~kg/m$^2$. 
\begin{figure}
    \begin{center}
    \includegraphics[width=0.7\textwidth]{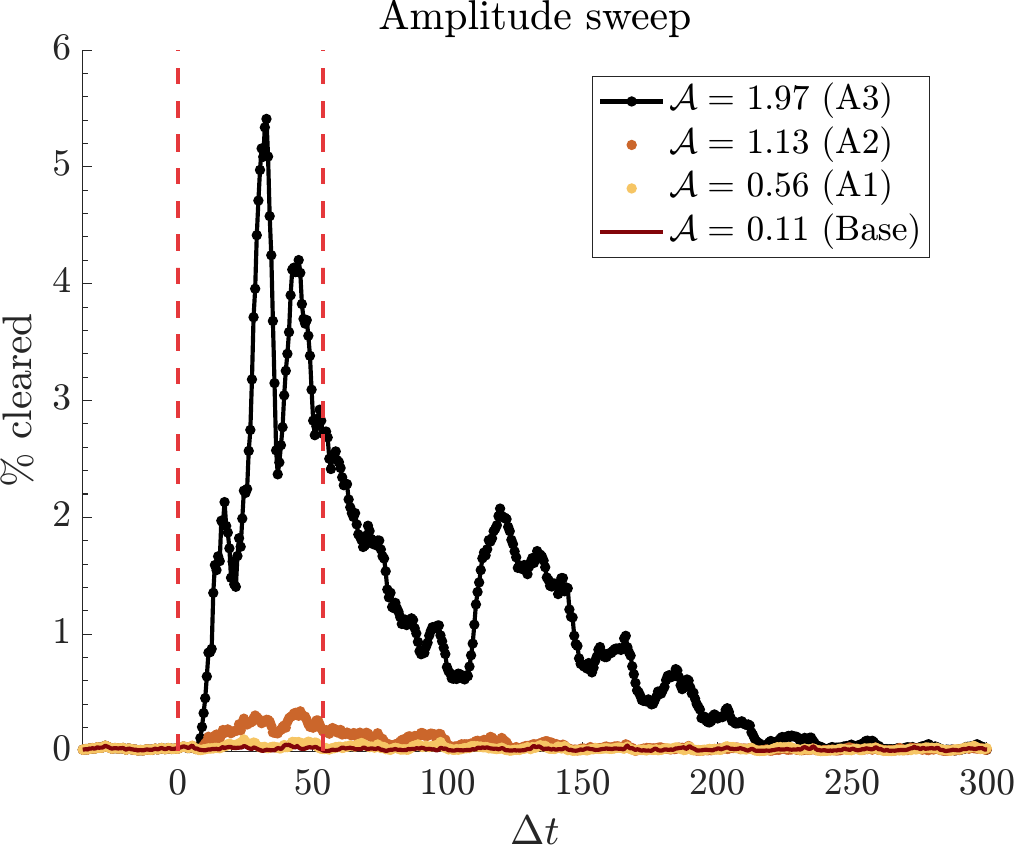}
    \caption{Total cloud cleared percentage as a function of time since the forcing onset for Base, A1, A2, and A3. The red dashed lines denote the time window in which the forcings were active. \label{fig: Asweep}}
    \end{center}
\end{figure} Fig. \ref{fig: Asweep} shows the time trace of cleared cloud for the four cases in which the amplitude was varied. Here, $\Delta t$ denotes the time relative to the forcing onset normalized by $t_s$ ($\Delta t = 0$); a negative $\Delta t$ refers to the RCE period (when the cloud deck coverage was 100\%). The observationally derived forcing (Base, $\mathcal{A}=0.11$) led to no change in the cloud deck, as did a forcing whose amplitude was five times larger. Some weak signature of breakup is observed for $\mathcal{A} = 1.13$, which supports the nondimensionalization choice to derive Eq. \ref{eq: nondimMom}: an $O(1)$ term is necessary to overcome the RCE balance. Increasing the amplitude further to $\mathcal{A} = 1.97$ leads to a pronounced signal and a peak breakup of nearly $6\%$. Curiously for A3, there is a second peak within 5 time units of the first peak at $\Delta t = 33$ before the recovery of the cloud deck after the forcing ceased. This second peak is not correlated with the zero-crossing of the forcing nor with the time trace of entrainment velocity, suggesting the behavior is likely attributable to additional nonlinear effects between the forcing and STBL \citep{ARB25}. All four cases exhibit an adjustment back to the stationary state by $\Delta t = 300$, though the STBL under the largest amplitude forcing experiences another slight uptick in clearing near $\Delta t = 120$. After this uptick, there is a decaying oscillation-like behavior with a period of approximately $(L_x/\ov{z}_i) \sec(v_o/u_o)$, which is the ``recirculation'' time for a cloud parcel to re-enter the domain due to the periodic boundary conditions \citep{ARB25}. This signature is characteristic of the other experiments as well.  

Given that the A3 case exhibited the strongest breakup, we will fix this amplitude (and thus the wave energy) when analyzing the effect of the forcing period and location. 
\begin{figure}
    \begin{center}
    \includegraphics[width=1\textwidth]{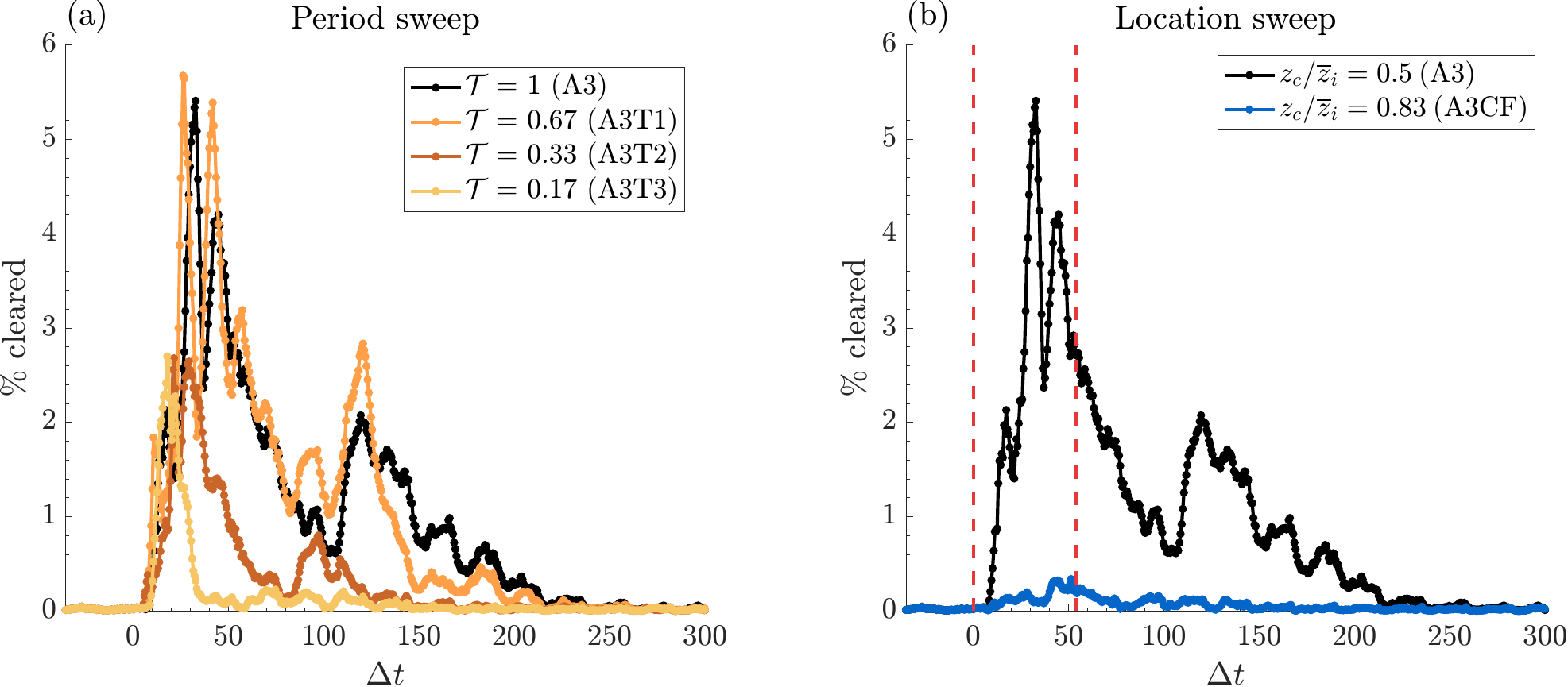}
    \caption{Total cloud cleared percentage for $\mathcal{A}=1.97$ varying (a) period and (b) vertical center location of the forcing. \label{fig: TA0p175}}
    \end{center}
\end{figure} Fig. \ref{fig: TA0p175} depicts the overall clearing at this amplitude level as a function of period ($\mathcal{T} = 1/[\omega/\omega_{ref}]$) and location. While Fig. \ref{fig: TA0p175}a suggests that longer period waves lead to more clearing, this is largely a function of the longer duration of forcing for those cases (since each case is only forced for one integer period). All four cases exhibit a similar trend in LWP reduction upon the onset of forcing at $\Delta t = 0$, but the RCE recovery for the shorter period forcings ($\mathcal{T} = 0.17,0.33$) is faster than their longer period counterparts ($\mathcal{T} = 0.67,1$). Curiously, A3T1 ($\mathcal{T} = 0.67$) exhibits the same level of breakup as A3 and a more pronounced ``rebound'' again near $t=120$ before settling back to the full cloud deck. 

Another forcing sweep member of interest is the vertical location over which the forcing is active. A3CF featured a centered, constrained forcing within the cloud layer (i.e., region of nonzero $q_c$ in Fig. \ref{fig: RCE}b), subjecting the forcing near the strongest stratification in the STBL with $N_m\sim10^{-2}$~s$^{-1}$ (not shown). However, Fig. \ref{fig: TA0p175}b illustrates that despite A3CF being more suited to sustaining generated waves, the locality of the forcing attenuated its clearing ability, similar to the level observed in A2.  

\begin{figure}
    \begin{center}
    \includegraphics[width=0.7\textwidth]{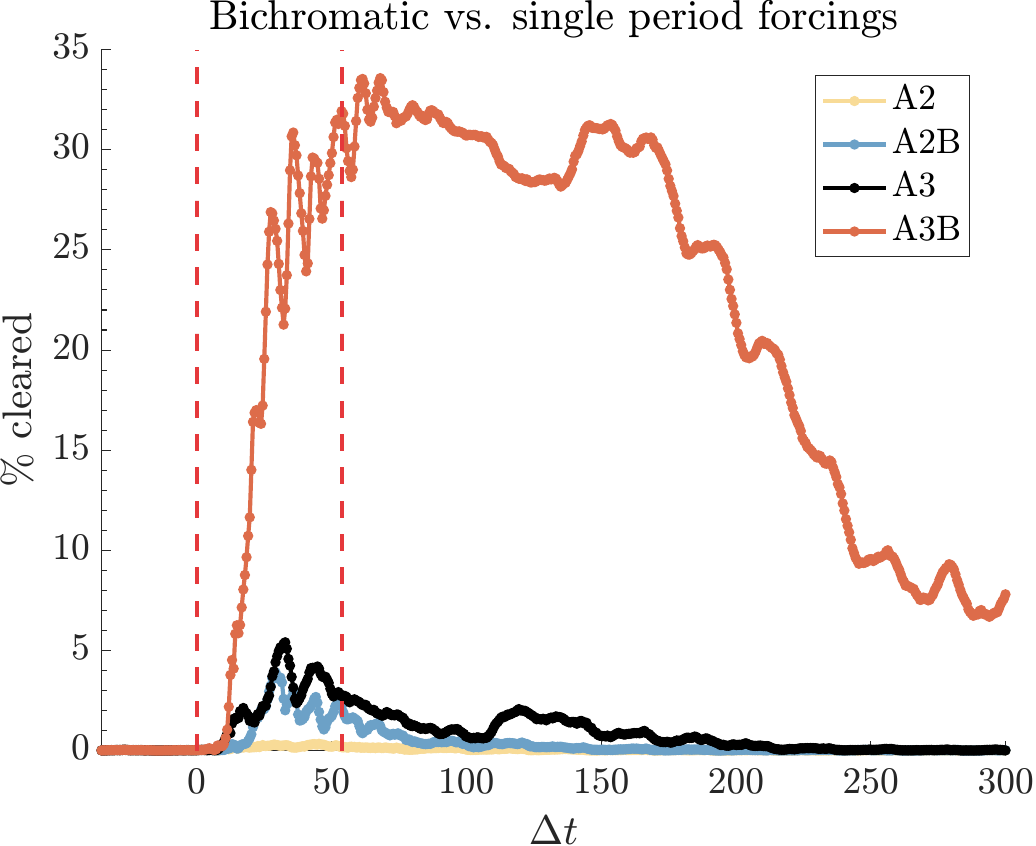}
    \caption{Same as Figs. \ref{fig: Asweep}-\ref{fig: TA0p175}, but for the bichromatic forcing cases (A2B, A3B) compared against the monochromatic experiments (A2, A3). Again, the vertical red dashed lines indicate the forcing active window. \label{fig: bi}}
    \end{center}
\end{figure}

Lastly, the impact of a forcing with the superposition of two frequencies was explored. This is seen in Fig. \ref{fig: bi}. 
 Each bichromatic forcing was composed of the longest and shortest period wave tested in this study. For A2B, addition of this second frequency led to similar clearing levels as A3 alone. By contrast, A3B led to dramatic clearing of the deck with levels exceeding 30\%. The underlying high frequency of the forcing is active for six cycles, augmenting the single period of the low frequency component thereby enhancing breakup. 

\begin{figure}
    \begin{center}
    \includegraphics[width=1\textwidth]{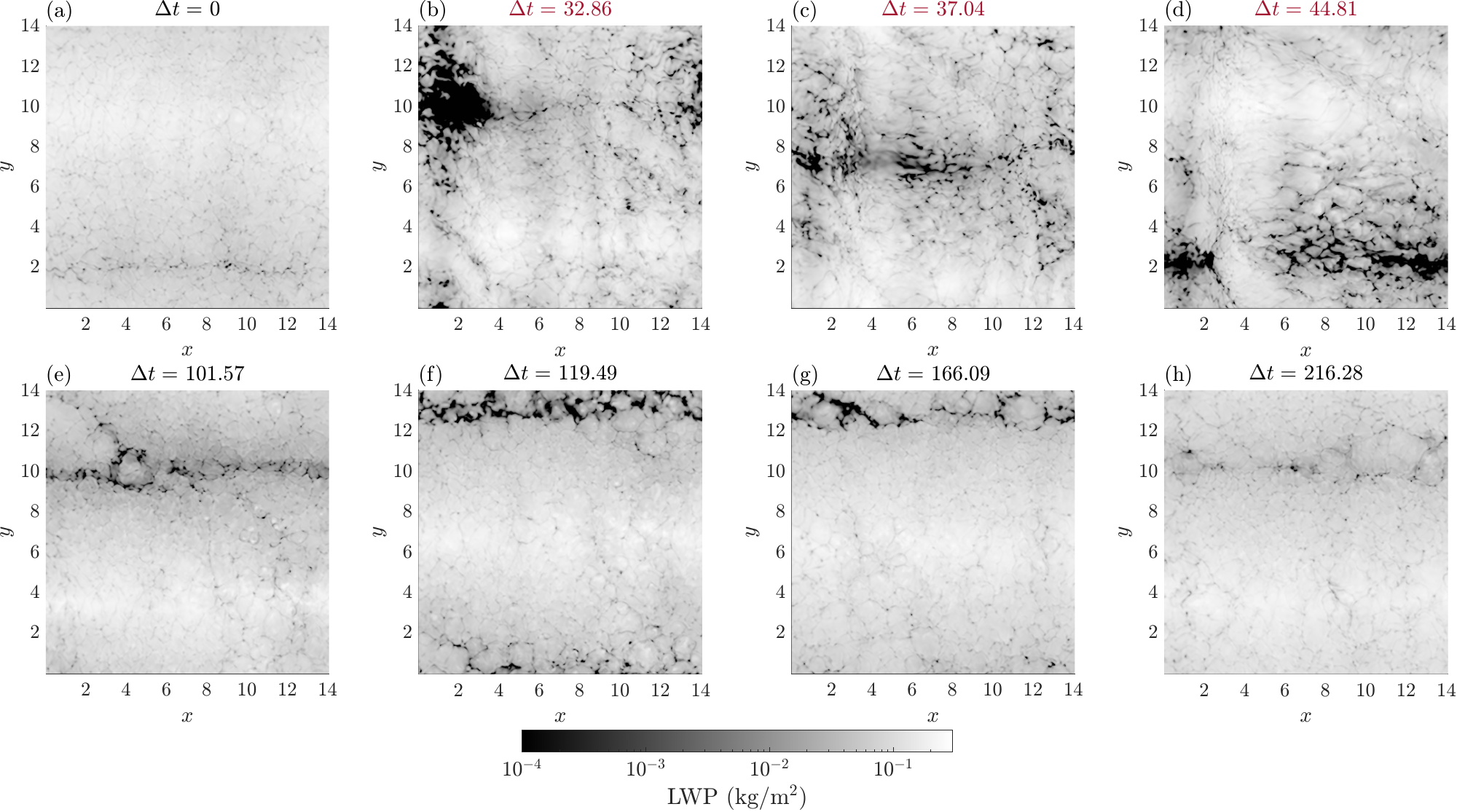}
    \caption{Snapshots of LWP through time for case A3. Red plot titles indicate active forcing. Panel b) is at the max clearing time and f) is at the peak ``rebound'' point after the forcing ceased, as seen in Fig. \ref{fig: Asweep}. \label{fig: LWPA0p175}}
    \end{center}
\end{figure}

\begin{figure}
    \begin{center}
    \includegraphics[width=1\textwidth]{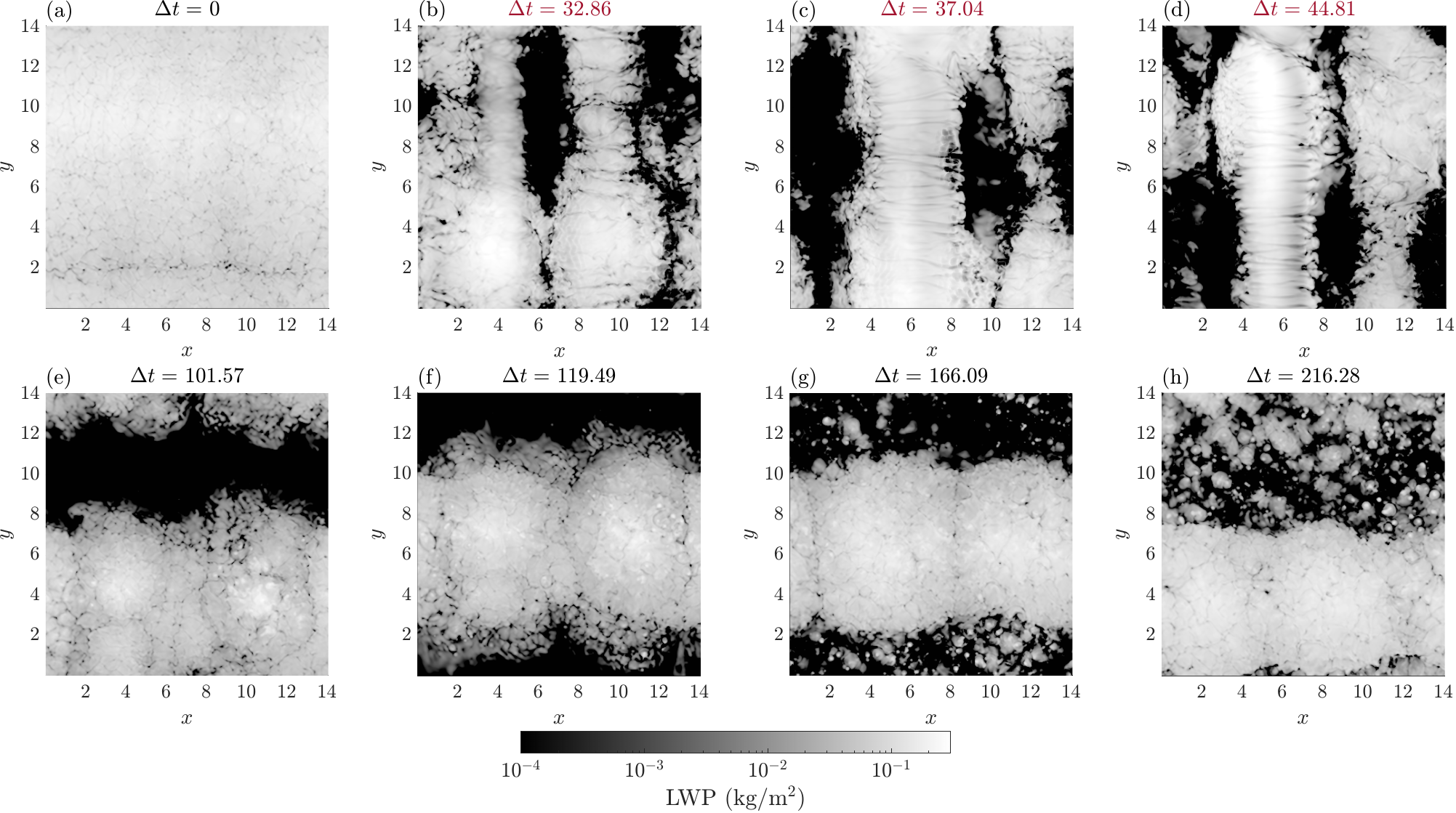}
    \caption{LWP snapshots for case A3B at the same times as Fig. \ref{fig: LWPA0p175}. Red plot titles indicate active forcing.  \label{fig: LWPA0p175Bi}}
    \end{center}
\end{figure}

Figs. \ref{fig: LWPA0p175}-\ref{fig: LWPA0p175Bi} illustrate the full liquid water path field through time. For A3, the panels show that open cellular structures are initially clustered together. As time progresses, these open cells are advected to the southeast by the mean winds forming a horizontal line of POC \citep{ARB25}. In the bichromatic forcing case A3B, the area of depleted LWP is far larger than A3 with the cleared areas following the peaks/troughs of the excited waves during the forcing period (panels b-d in Fig. \ref{fig: LWPA0p175Bi}). After the forcing ceased, there is a persistent thick band of cloud interspersed between rectangular patches of clear sky. 

\subsection{Energetics}
To better understand how the turbulent state is modulated by the forcing, we analyze budgets of the turbulent kinetic energy (TKE) equation, 

\begin{multline}\label{eq: tkeunravel}
    \ub[1pt]{\f{\pd}{\pd t}\bigg(\f{1}{2}\ov{u'_i u'_i}\bigg)}_\text{\clap{$\pd k/\pd t$~}} = \ub[1pt]{-\overline{u_i'u_j'}\f{\pd \ov{u}_i}{\pd x_j}}_\text{\clap{$\mathcal{P}$~}}\ub[1pt]{-\mathrm{Ja}_m\ov{u'_i\theta_\rho\f{\partial \pi'}{\partial x_i}}}_\text{\clap{$\Pi$~}} \ub[1pt]{-\ov{u}_j\f{\pd }{\pd x_j}\bigg(\f{1}{2}\ov{u_i'u_i'}\bigg)-\ov{u'_j\f{\pd}{\pd x_j}\bigg(\f{1}{2}u'_iu'_i\bigg)}}_\text{\clap{$\tau$~}} + \ub[1pt]{\f{1}{\mathrm{Fr}^2} \ov{u'_i\theta'_\rho}\delta_{i3}}_\text{\clap{$\mathcal{B}$~}} \\ + \f{\ov{z}_i}{U^2}\big(\ub[1pt]{\ov{u'_iT'^*_i} + \ov{u'_iN'^*_i}}_\text{\clap{$\varepsilon$~}} +\ub[1pt]{\ov{u'_i\mathcal{F}'^*}\delta_{i3}}_\text{\clap{$\Lambda$~}}\big). 
\end{multline} The terms on the right-hand-side are production ($\mathcal{P}$), pressure correlation ($\Pi$), combined mean and turbulent transport of TKE ($\tau$), buoyancy generation/destruction ($\mathcal{B}$), dissipation ($\epsilon$), and forcing injection ($\Lambda$). Note that we neglect the TKE due to subsidence, as it is quite small relative to the other terms \citep{ARB25}. The TKE budget also closes as expected: see Appendix C for a more detailed discussion.  
\begin{figure}
    \begin{center}
    \includegraphics[width=0.8\textwidth]{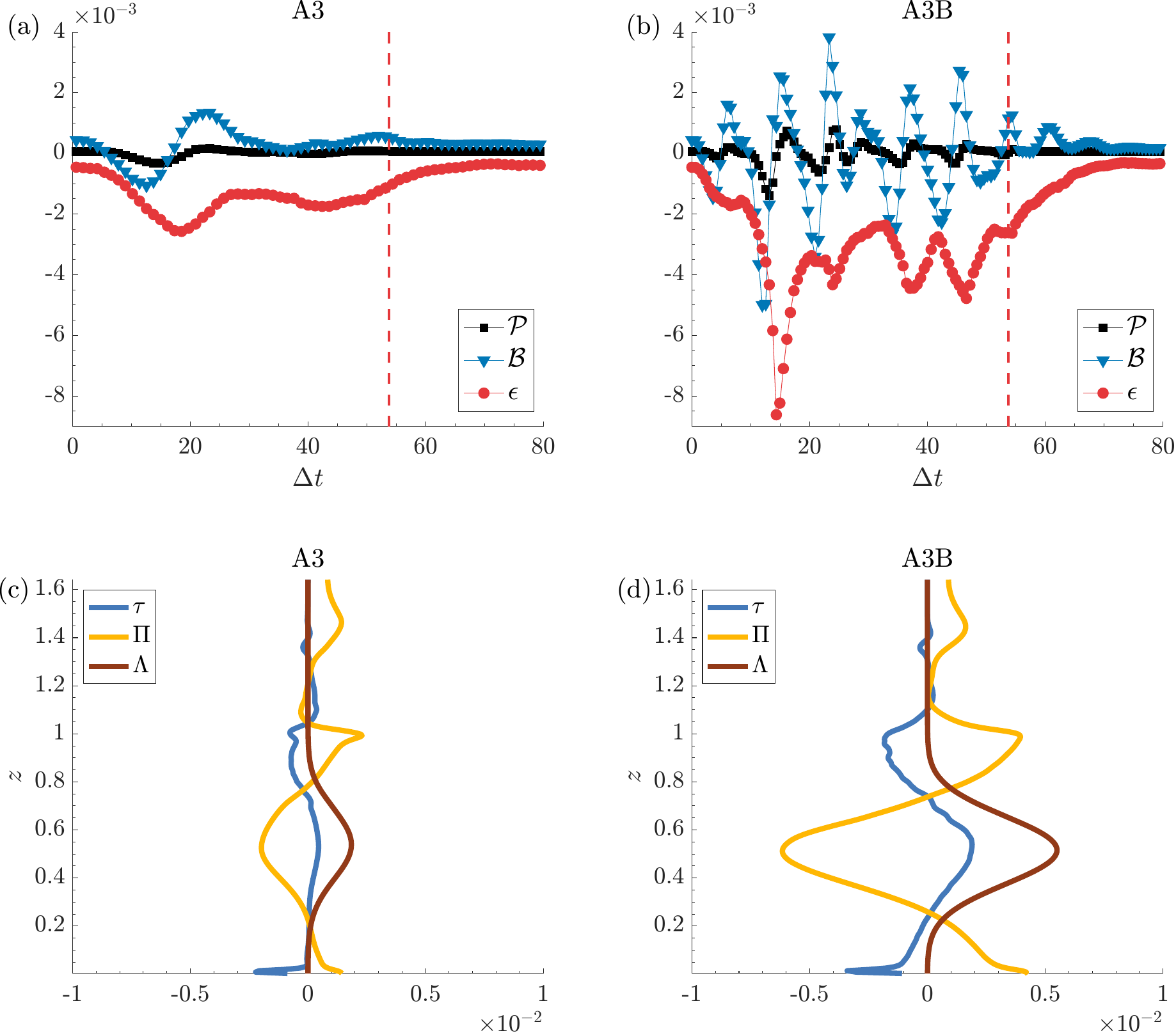}
    \caption{Dimensionless TKE budget terms as a function of time upon volume averaging (a,b) and $z$ averaging over three forcing periods. Panels (a,c) represent case A3 and (b,d) for A3B. Again, the red dashed line denotes the end of the forcing window in (a,b). Recall that $\mathcal{P}$ is shear production, $\mathcal{B}$ is buoyant production,  $\varepsilon$ is dissipation, $\tau$ is turbulent transport, $\Pi$ is the pressure correlation, and $\Lambda$ is the TKE due to the forcing. \label{fig: TKEbudgetA3s}}
    \end{center}
\end{figure} Fig. \ref{fig: TKEbudgetA3s} shows the various budget terms from Eq. \ref{eq: tkeunravel} evolving through time and as a function of the vertical for cases A3 (left panels) and A3B (right panels). In terms of the temporal plots, the shape of the buoyancy curves has the same structure as the imposed forcing; Fig. \ref{fig: TKEbudgetA3s}a) illustrates $\mathcal{B}$ undergoing one periodic oscillation while Fig. \ref{fig: TKEbudgetA3s}b) shows six. For A3, the buoyancy term plays a dual role in draining TKE to the mean kinetic energy for the first 20 time units, while energizing the turbulence in the latter part of the active forcing window \citep{ARB25}. Similar behavior is observed for A3B except that the generation-destruction pattern follows that of the high-frequency component of the forcing. Turbulent dissipation in both cases lags the buoyancy term. Furthermore, the TKE production is close to zero for A3 and no more than half of $\mathcal{B}$ for A3B with $\mathcal{P} < 0$ corresponding to the troughs of $\mathcal{B}$. In the latter case, the magnitude of $\mathcal{P}$ is primarily due to vertical gradients in $u$ and $v$ near the surface; mean wind shear in the horizontal directions is not a factor.  

How TKE is distributed throughout the boundary layer can be understood by analyzing the transport and pressure correlation terms. These are shown in Fig. \ref{fig: TKEbudgetA3s}c)-d) for A3 and A3B, respectively. Peak injection of TKE via the forcing at $z\sim0.5$ is mediated by the pressure correlation transferring energy back to the mean flow and returning that energy to the turbulence near the inversion, $z\sim1$. The combined transport term in both cases is largely opposite in sign to $\Pi$ \citep{ARB25}. 

Lastly, since the gravity wave forcing is introduced as a source term solely in the vertical direction, it is insightful to further characterize the anisotropy of the flow. This is quantified by the Reynolds stress anisotropy tensor

\begin{equation}\label{eq: bij}
    b_{ij} = \f{\ov{u_i'u_j'}}{\ov{u_k'u_k'}}-\f{1}{3}\delta_{ij}. 
\end{equation}  

\begin{figure}
    \begin{center}
    \includegraphics[width=0.7\textwidth]{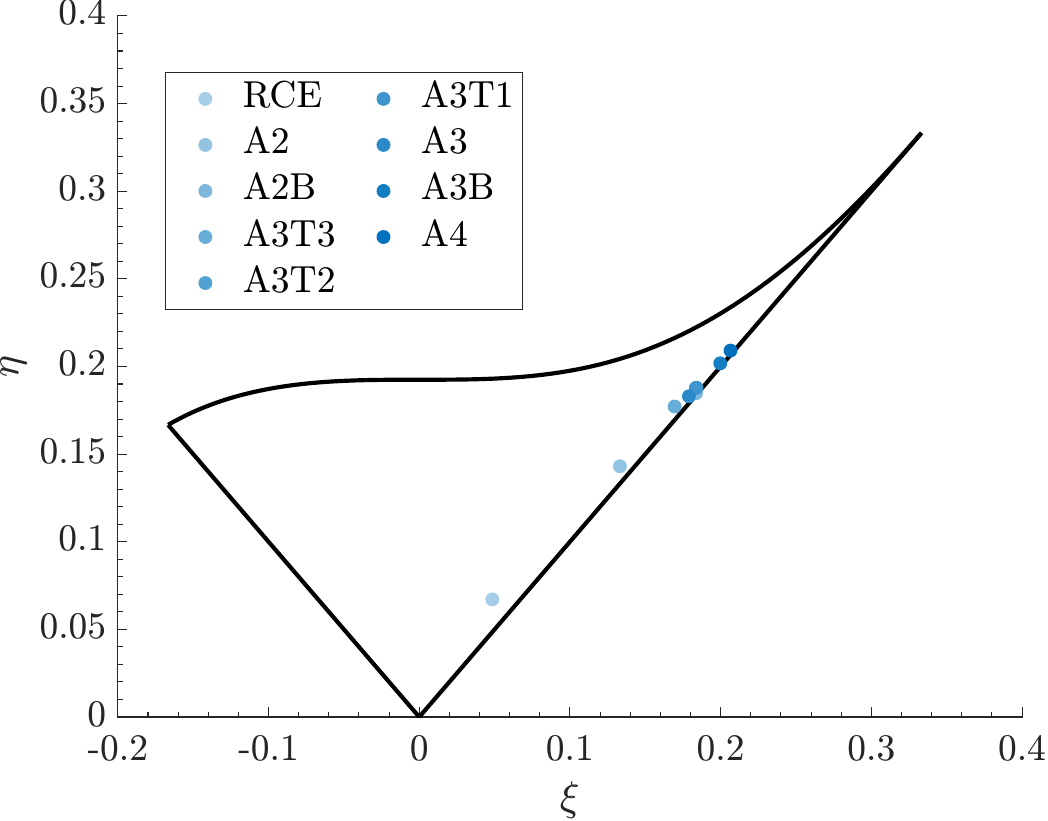}
    \caption{Lumley triangle diagram for all experiments. The anisotropy tensor is volume and time averaged over each experiment's respective forcing period. \label{fig: lumley}}
    \end{center}
\end{figure}

Fig. \ref{fig: lumley} is a Lumley triangle visualization where $\xi$ and $\eta$ represent invariants of $b_{ij}$ which can be expressed in terms of its eigenvalues \citep{Pope}.  The $(0,0)$ point represents the isotropic limit, whereas the right corner indicates that one  diagonal component of the Reynolds stress is dominant \citep{yi22}. Conversely, the left corner marks where two components of $b_{ii}$ outweigh the third. The black lines of the ``triangle'' bound the possible turbulent states, providing a unifying way to compare the turbulent character of each type of STBL forcing and how the values of the invariants relate to breakup \citep{ARB25}. For each of the experiments, we perform a volume average and a temporal average of Eq. \ref{eq: bij} over one forced period and then compute $\xi$ and $\eta$. The various experiments are colored in shades of blue with darker colors representing cases that experienced greater reductions in LWP. The RCE case sits near the bottom of the axisymmetric $\eta = +\xi$ line. As the level of breakup increases from lower amplitude to higher amplitude experiments, their $(\xi,\eta)$ coordinate translates toward the one-component ``rodlike'' limit. Analyzing the components of $b_{ii}$ as a function of $z$, the dominant component during the forcing period is $\ov{u'u'}$ (not shown). This leads to an attenuation in $\ov{v'v'}$ and $\ov{w'w'}$ on account of $b_{ij}$ being traceless, suggesting that the forcing primarily deposits TKE in the streamwise fluxes. 

\begin{figure}
    \begin{center}
    \includegraphics[width=1\textwidth]{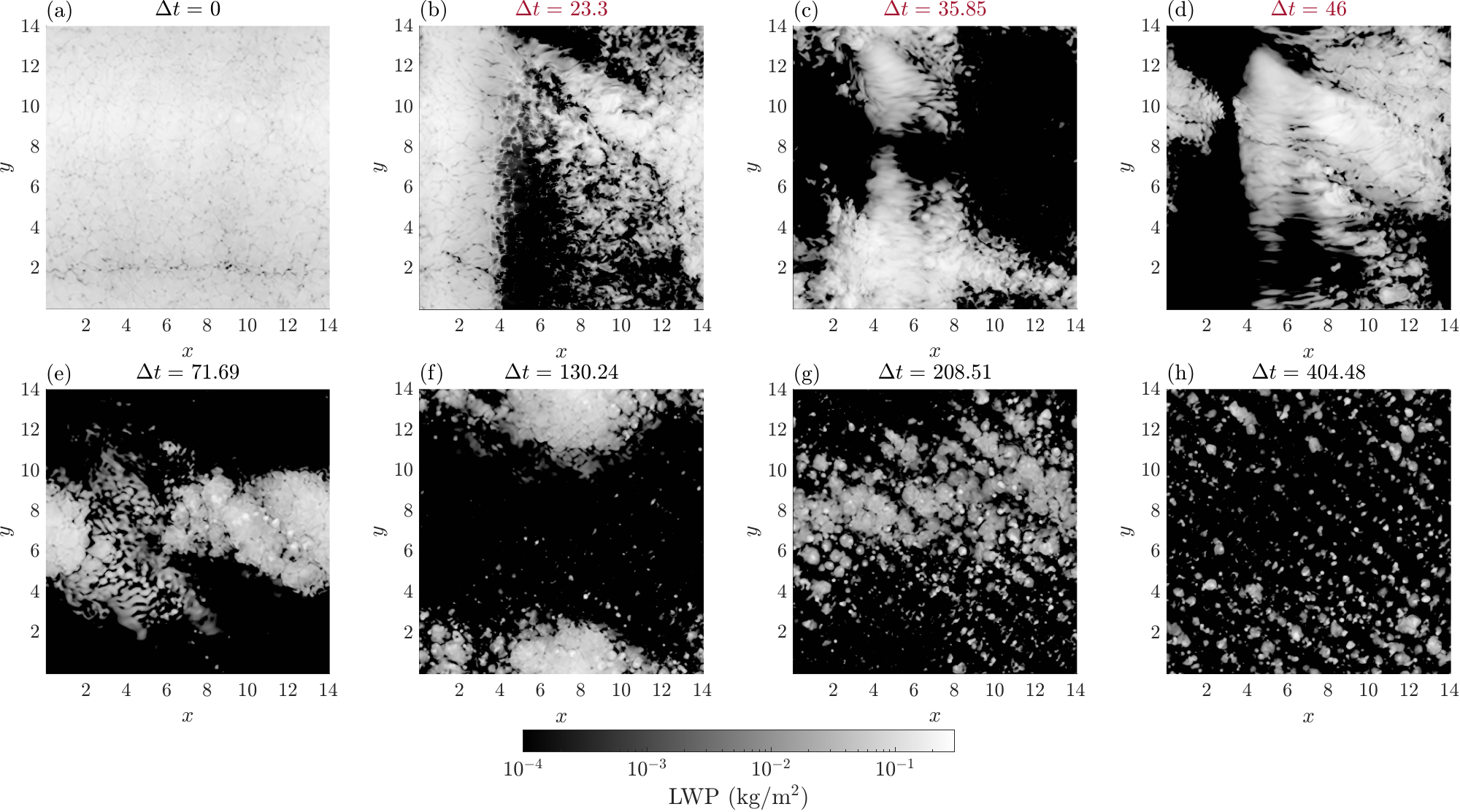}
    \caption{LWP snapshots for case A4. Red plot titles indicate active forcing.  \label{fig: LWPA0p25}}
    \end{center}
\end{figure}

\begin{figure}
    \begin{center}
    \includegraphics[width=1\textwidth]{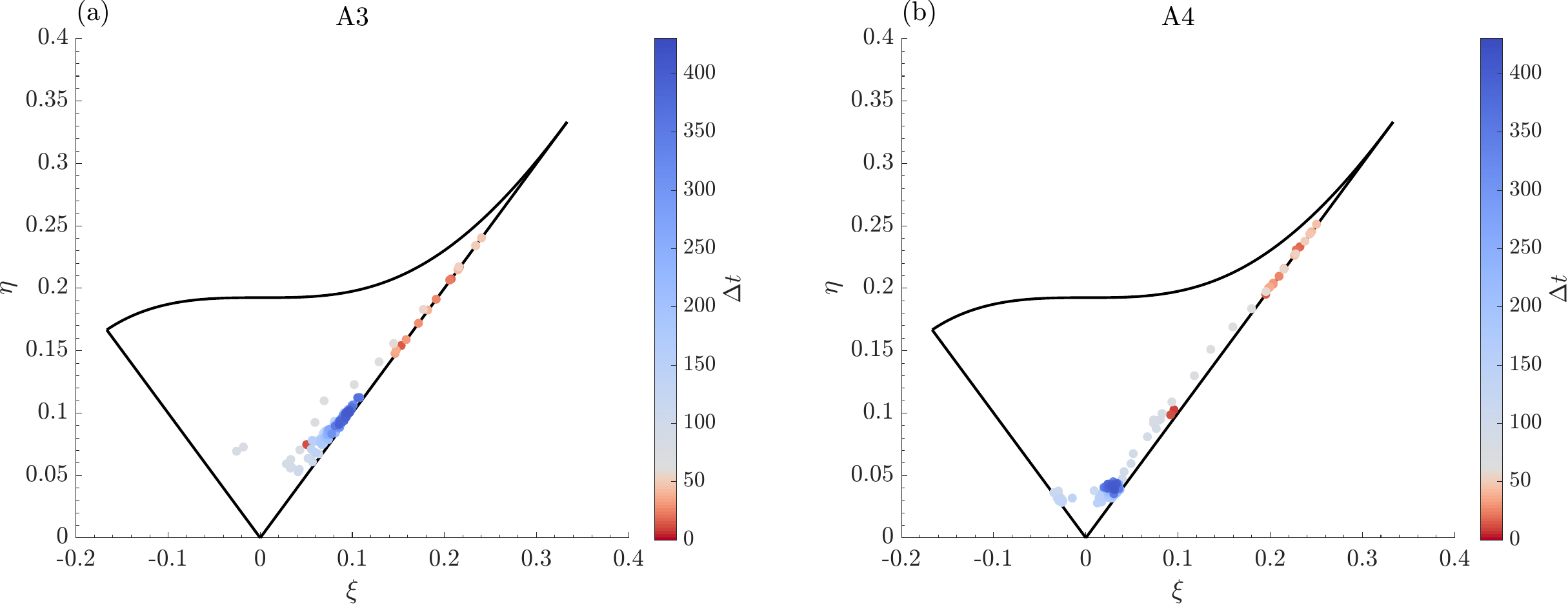}
    \caption{Lumley triangle diagram for case a) A3 and b) A4. Colored markers show progression through time. Red markers indicate active forcing time. \label{fig: lumleyA3A4}}
    \end{center}
\end{figure}

Fig. \ref{fig: LWPA0p25} shows one such case (A4) where the STBL does not transition back to RCE. Not only does the cloud deck dramatically clear during active forcing seen in Fig. \ref{fig: LWPA0p25}(b-d), this is sustained after the forcing ceases. This suggests that A4's Lumley triangle coordinate is an upper bound of anisotropy for this type of forcing. Fig. \ref{fig: lumleyA3A4} shows the full time progression of the invariants for A3 and A4. As forcing ensues for both cases, the $(\eta,\xi)$ coordinate again approaches the top-right rodlike limit of $b_{ij}$. After excitation ceases around $t\sim50$, the A3 stress state converges back to the RCE point \citep{ARB25}. Not only does the cloud deck return to a closed cellular regime, but the mean turbulent state at the final time of A3 has no knowledge of the forced waves excited at an earlier time. In the case of A4 after forcing stops, its stress state appears to traverse towards the $\eta = -\xi$ disklike leg, where $\ov{u'u'}$ transfers its energy to briefly reenergize $\ov{v'v'}$ and $\ov{w'w'}$, before settling again near the rodlike leg. This behavior exemplifies the change in the turbulent state relative to its RCE counterpart in this largely cloud-free regime, occupying a final state closer to the isotropic limit.

\subsection{Amplitude scaling relationships}
\begin{figure}
    \begin{center}
        \includegraphics[width=0.7\linewidth]{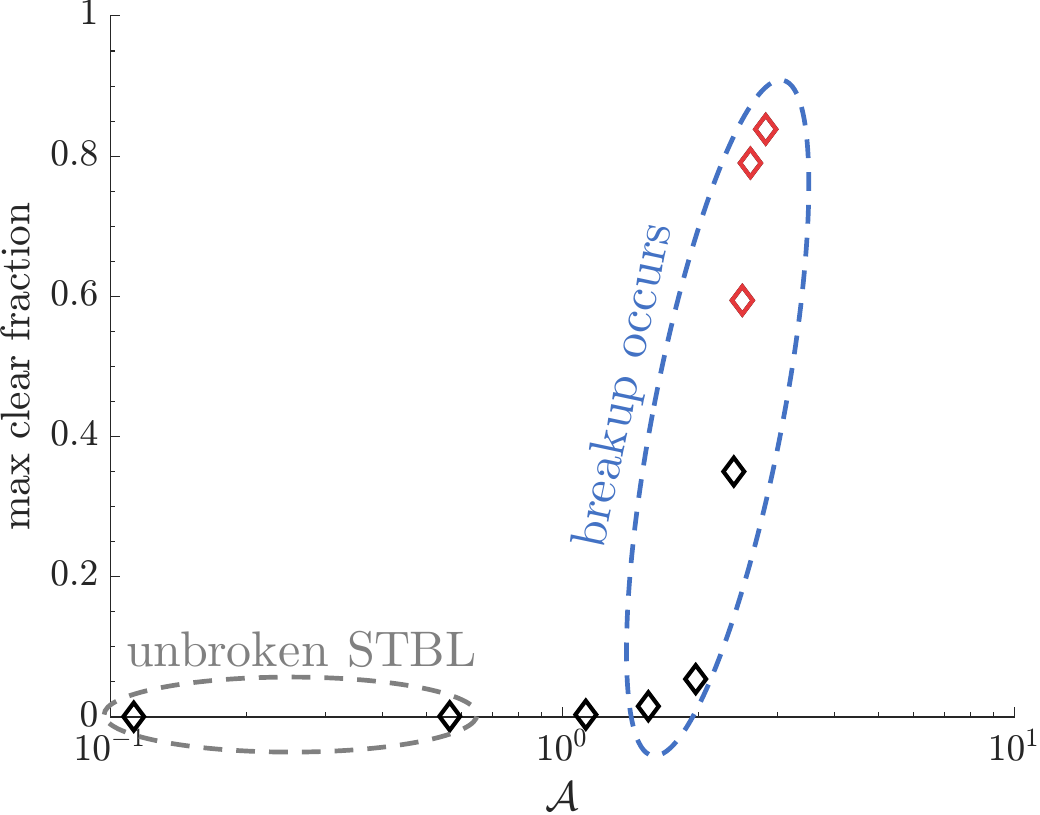}
        \caption{Maximum cloud clearing fraction versus forcing amplitude. The gray oval highlights simulations where the STBL remains unbroken while the blue oval shows the cases where breakup occurs. The red markers illustrate forcings that transitioned the STBL to a quasi-permanent broken regime.}
        \label{fig: Ascale}
    \end{center}
\end{figure}

As seen above in \S \ref{sec:results}b, increasing the forcing wave energy leads to more clearing in a transient sense. A question that remains is whether there exists a critical minimum forcing amplitude that pushes the STBL away from RCE in a quasi-permanent manner \citep{ARB25}. Fig. \ref{fig: Ascale} shows the maximum fraction of cloud cleared for each forcing amplitude at any point during the simulation. 

Here, there is a clear demarcation between unbroken versus broken STBL based on whether the forcing amplitude $\mathcal{A} < 1$ or $\mathcal{A} > 1$, respectively. Again, Fig. \ref{fig: Ascale} lends support for the nondimensionalization in Eq. \ref{eq: nondimMom}. The three highest amplitude cases are colored in red to indicate simulations where a patchy cloud deck was sustained. It is apparent that $\mathcal{A} = 2.5$ seems to be a critical amplitude that separates an STBL that transitions back to stationarity from one that becomes permanently broken. 

We can further assess the impact of cleared cloud by integrating Fig. \ref{fig: Asweep} through time

\begin{equation}
    R \propto \int_{t_i}^{t_f} \% \text{ cleared } dt
\end{equation} where $R$ is a proxy for the disruption in cloud radiative cooling when breakup occurs. From $R$, we can also extract a characteristic clearing time scale $t_c$ based upon the maximum fraction of cloud cleared during each run, 

\begin{equation}
    t_c = \f{R}{\%_{max}}. 
\end{equation}

\begin{figure}
    \begin{center}
        \includegraphics[width=1\linewidth]{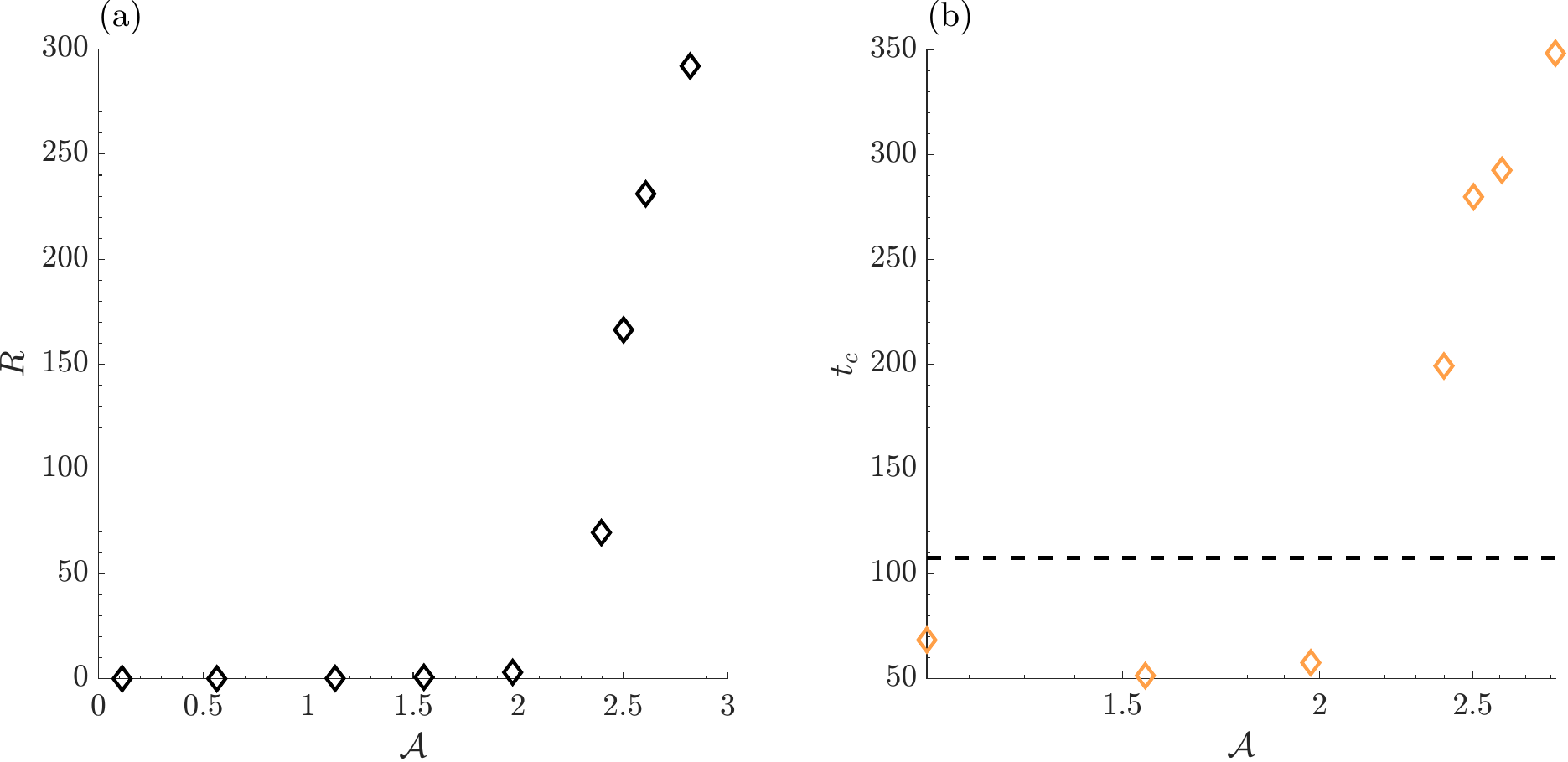}
        \caption{a) Radiative cooling anomaly and b) characteristic clearing time scale versus forcing amplitude. The black dashed line in panel b indicates the imposed longwave cooling time scale $\tau_c$ in Eq. \ref{eq: Qdot} nondimensionalized by $t_s$.}
        \label{fig: tcRscale}
    \end{center}
\end{figure} These parameters are shown as a function of forcing amplitude in Fig. \ref{fig: tcRscale}. Panel a) shows that for $\mathcal{A} < 2$, this radiative cooling anomaly metric is essentially zero, which is as expected since these cases transitioned back to RCE given sufficient time after forcing ceased. Only for $\mathcal{A} \ge 2$ do we begin to see significant values of $R$ with the largest amplitude having a nearly 6 times larger disruption of radiative cooling than the critical $\mathcal{A} = 2.5$ case. Panel b) further informs the separation of STBL regimes in terms of the imposed longwave cooling time scale $\tau_c$ in Eq. \ref{eq: Qdot}. Having a long characteristic clearing time is emblematic of clearing that persists, which is again seen for the $\mathcal{A} > 2$. For $\mathcal{A} < 2$, the shorter $t_c$ over which clearing occurs is less than $\tau_c$, indicating that eventual return to RCE is facilitated by the radiative cooling model reestablishing cloud liquid water near the inversion. Larger $t_c$ in the corresponding cases indicates that this time scale dominates over $\tau_c$ and thus leads to a broken, patchy regime even after forcing ceases. This trend in $t_c$ is consistent with the bichromatic simulations as well (not shown). 

\begin{figure}
    \begin{center}
        \includegraphics[width=0.7\linewidth]{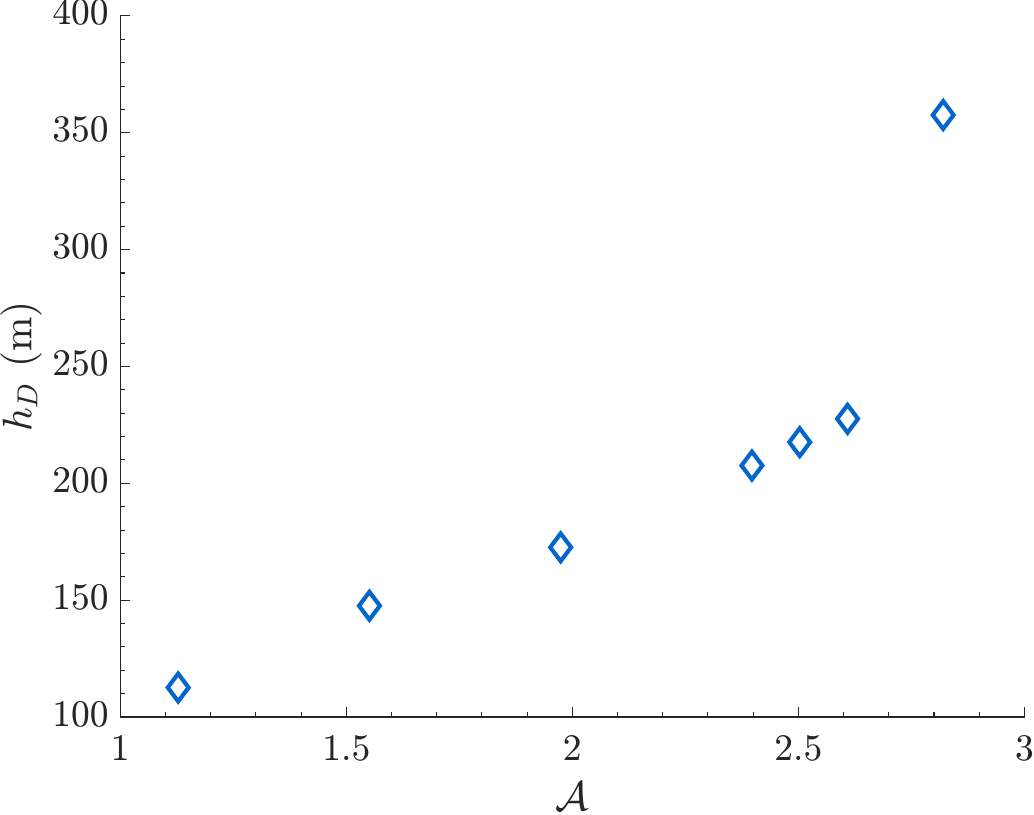}
        \caption{Maximum cloud displacement height as function of forcing amplitude.}
        \label{fig: hD}
    \end{center}
\end{figure}

\begin{figure}
    \begin{center}
        \includegraphics[width=1\linewidth]{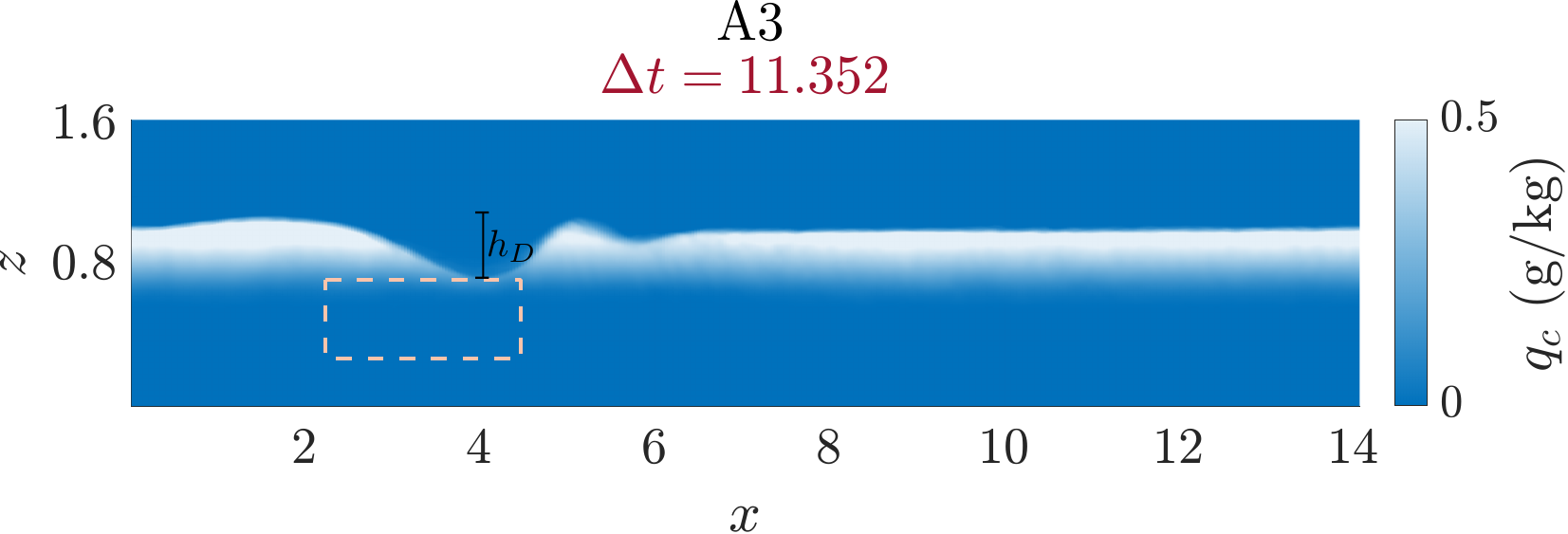}
        \caption{Cloud condensate specific humidity averaged in $y$ for case A3 at the time of maximum compression, labeled $h_D$. The dashed rectangle shows the approximate forcing region.}
        \label{fig: qcxz}
    \end{center}
\end{figure}

Lastly, the forcing influences the vertical structure of the cloud deck by inducing a train of streamwise-propagating waves. Supplemental movies S1-S2 show the excited gravity wave modes in the $q_c$ field for A3 and A4, which cause the cloud deck to compress/thin. The vertical cloud-top compression is analogous to the level of dry air entrainment promoted by the forcing. Fig. \ref{fig: hD} shows the cloud displacement level ($h_D$) as a function of forcing amplitude. The displacement is the maximum  distance the cloud top traverses downwards relative to the RCE state over all $x$ and during the negative phase of the forcing; $h_D$ is indicated in Fig. \ref{fig: qcxz} for case A3. As forcing amplitude increases, $h_D$ increases in a near linear manner with exception of the highest amplitude value. For the patchy STBLs ($\mathcal{A}\geq2.5$), these cases had downward displacement figures ranging from $200-400$~m, corresponding to 22-44\% of the RCE-average inversion height. More work is necessary to understand this scaling from a reduced order perspective. 

\section{Summary and discussion} \label{sec:conclusion}
We have explored the effect of a gravity wave forcing on the stratocumulus-topped boundary layer with LES. The forcing took the form of a plane wave and acted as a source term for the vertical momentum equation. In the limit of a dry, shear-free configuration, the forcing recovers the linear dispersion relation of internal gravity waves \citep{ARB25}. (For further information, see Appendix A.) 

In order to ascertain the impact of the forcing itself, we constructed a novel radiative-convective equilibrium framework for the STBL by introducing a newtonian relaxation term in the area above the cloud that represents free-troposphere cooling. The degree of stationarity is measured by the equivalent potential temperature. Initializing the STBL with field-campaign derived profiles from \cite{Stevens2005LES} leads to the onset of RCE after 125 hours of model spin-up \citep{ARB25}. 

The characteristics of a nominal forcing were estimated from satellite measurements of these wave packets in the southeastern Pacific \citep{Connolly2013,Allen2013}. From this baseline, we varied key parameters of the forcing such as its amplitude, period, locality, and chromaticity. Only forcings whose amplitude exceeded unity led to any large-scale change in the LWP which lends support for the proposed nondimensionalization in Eq. \ref{eq: nondimMom} \citep{ARB25}.

Upon fixing the amplitude above the unity breakup threshold, longer period forcings led to more breakup by virtue of each experiment having an active forcing window of one integer period, though A3T1 had the same level of clearing as A3 despite forcing for less time. The cases that led to the strongest breakup relative to a lower input wave energy were the bichromatic forcings; nonlinear interaction between the two forced frequencies promoted larger areas of clear sky that sustained for longer as well. Snapshots of the LWP illustrated large areas of clear skies that persisted through time, suggesting that the entire gravity wave spectrum, if their energy is sufficient, can play a role in affecting mesoscale cloud structure. Further scaling analysis showed that forcing amplitudes larger than 2.5 led to a largely-cleared STBL even after the forcing ceased, suggesting that marked disruption of cloud-top cooling can occur via high-energy waves. Reduced order closures in global climate models should be refined to account for wave-induced cloud breakup and their dynamical interplay. Future observational studies are necessary to substantiate the wave amplitude range explored in this study as well as the choice of $t_s$ as the wave acceleration time scale. Measurements of gravity waves within the boundary layer remain a challenge as it is difficult to disentangle their properties from the background turbulence \citep{jia19}. However, studies have shown the existence of strong vertical velocity updrafts \citep{AGWw} and the potential for $O(1)$~km gravity wave displacement amplitudes in stratocumulus systems \citep{li2013AGWamp}. If the acceleration timescale of these AGWs is near $t_s$, this lends more credence to the realism of the amplitude sweep. 

Lastly, we looked at turbulent kinetic energy budgets as well as the degree of anisotropy across all the forced simulations. Analysis of the A3 case and its bichromatic counterpart revealed the dominance of the buoyancy generation term relative to shear production. The degree of anisotropy was also revealed to be a criterion that can be connected to breakup, as stronger cleared cases developed towards the one-component limit of the Lumley triangle, while patchy STBLs lie near the isotropic limit. Future studies with this novel STBL RCE-wave framework should also consider the impact of an interactive radiation model that includes shortwave radiation (i.e. sunlight). In addition, assessing the impact of the inherent Doppler shift by the STBL base wind profile could provide more physical insight into the temporal signature of the forcing. 

It would also be illuminating to further dissect the flow field into a mean ($\ov{u}_i$), fluctuating ($u'_i$), and wave components ($\Tilde{u}_i$) to better understand the interplay between the background and forcing modes superposed on the turbulent state. In order to construct $\Tilde{u}_i$, one must identify a singular, distinct period to define a phase average \citep{hussain1970mechanics}. Performing this decomposition may also prove insightful to understand the forcing amplitude threshold that separates a temporary broken regime to one where the clouds are persistently patchy \citep{ARB25}. 

\acknowledgments
A.B. acknowledges support from a National Science Foundation Graduate Research Fellowship (Grant Number DGE-1656518). The authors acknowledge computing resources provided by the Stanford Research and Computing Center and Dr. George Bryan for maintaining Cloud Model 1. A.B. also acknowledges fruitful discussions with Dr. Young R. Yi regarding the TKE budgets and Lumley triangle analyses presented in this study. Early versions of this work appear in the Center for Turbulence Research Annual Research Briefs (ARBs) written by A.B; duplicate portions of the ARB text are cited accordingly.

%
%
\datastatement
All files needed to reproduce the RCE, forcing, and gravity wave verification simulations are documented in the following Zenodo repository: \href{https://doi.org/10.5281/zenodo.18331430}{\textcolor{blue}{https://doi.org/10.5281/zenodo.18331430}}. The version of of CM1 used in this study is 21.0. Also included in the repository are the analysis codes, post-processed data, and supplementary movies. 

\begin{figure}
    \begin{center}
    \includegraphics[width=0.95\textwidth]{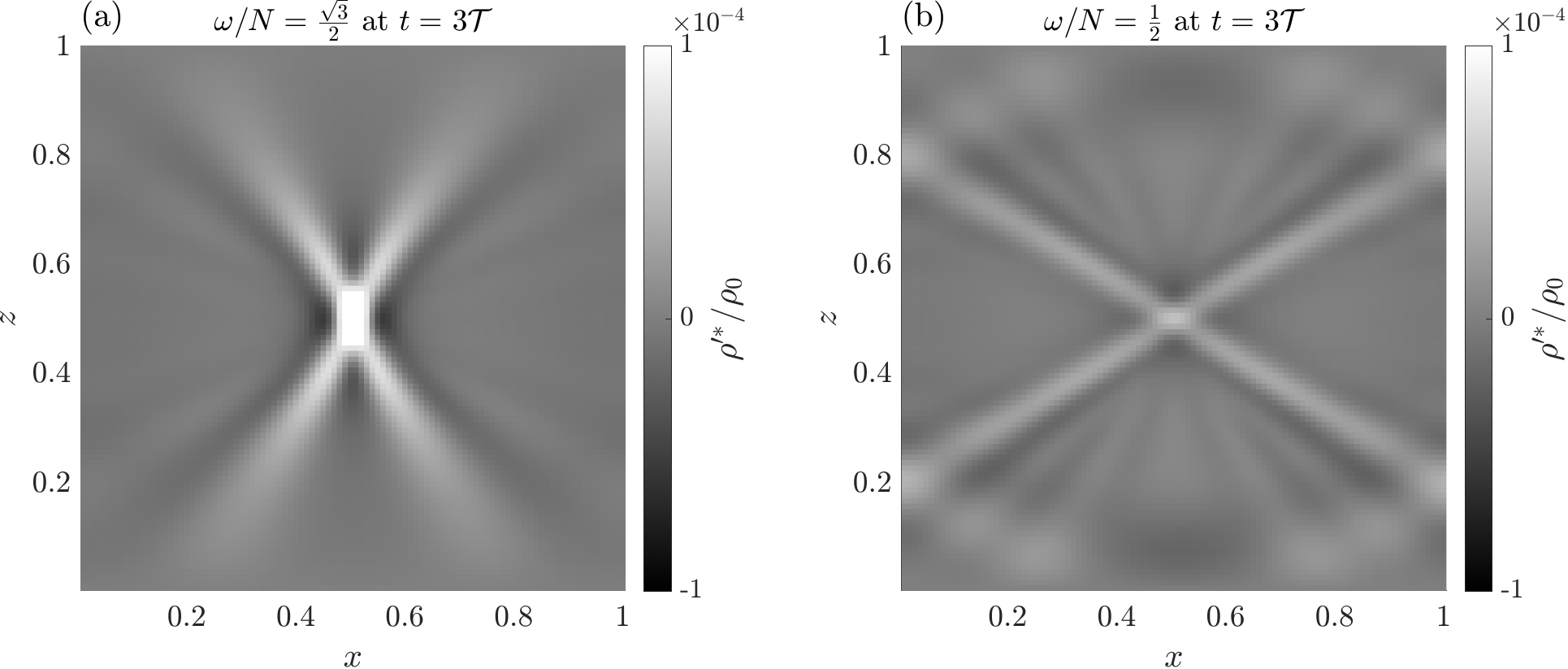}
    \caption{Dry, uniformly stratified setup with Eq. \ref{eq: GWforcing} imposed in the center of the domain. Dimensionless density perturbation contour of a) a forcing whose relative frequency yields a measured wave vector angle of 31.9° and b) 60.0° which agrees quite well to linear theory values of 30° and 60°, respectively. Both contours are shown after three periods of forcing. $N$ denotes the background buoyancy frequency. \label{fig: GWSt}}
    \end{center}
\end{figure}

\appendix[A]
\appendixtitle{Gravity wave forcing verification} 
In the absence of shear, inversion layers, and moisture, addition of Eq. \ref{eq: GWforcing} in a simply stratified setup should recover the linear dispersion relation for an internal gravity wave \citep{vallisBook}
\begin{equation}\label{eq: dispersion}
    \omega = N\cos{\alpha}
\end{equation} where $\alpha$ is the angle the wavevector makes with the horizontal and $N$ is the background stratification or buoyancy frequency. To test this, we ran a direct simulation of a dry, uniform $N = 10^{-2}$~s$^{-1}$ (a typical tropospheric value) setup mimicking a ``bobber'' in a wave tank. The imposed forcing had an amplitude of $10^{-3}$~m/s$^{2}$ and was applied in the center over an area approximately 0.5\% of the total domain size. Two distinct frequencies smaller than the background buoyancy frequency were chosen corresponding to a 30° and 60° wavevector angle. The phase lines, visualized with the density perturbations, corresponding to those cases are shown in Fig. \ref{fig: GWSt}. 
 Note that each case was forced for three integer periods. For further details, see the Zenodo repository in the \textit{Data Availability} section for the CM1 source code and input files. 

\begin{figure}
    \begin{center}
    \includegraphics[width=0.6\textwidth]{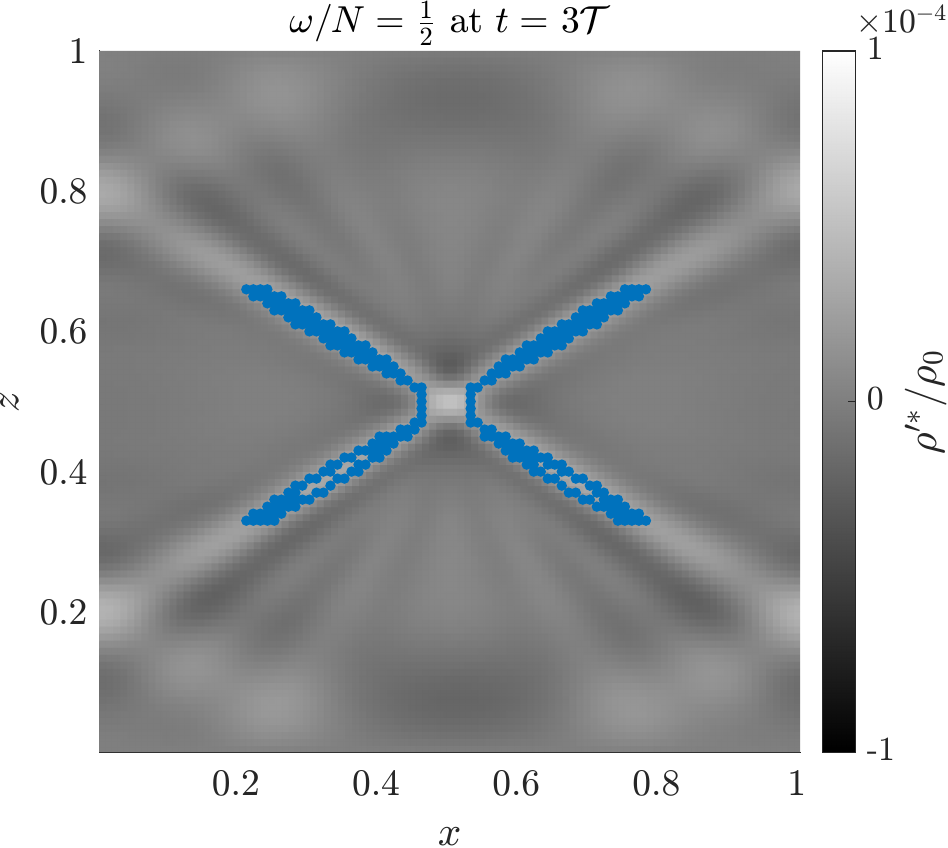}
    \caption{Density perturbation contour from Fig. \ref{fig: GWSt}b) overlaid with blue points that trace a constant contour value of $\rho'$ to compute the phase angle. \label{fig: stAangle}}
    \end{center}
\end{figure}

Fig. \ref{fig: stAangle} shows how the angle of the wavevector was extracted for the $\alpha = 60$° case. The blue dots signify a contour level that is 50\% of the maximum to isolate the x-shaped bands of the phase lines. Performing a linear regression on each branch and taking the complement of the arctangent of the slope yields the 60.0° estimate for the phase angle which agrees exactly with the dispersion relation. 

\appendix[B]
\appendixtitle{Background gravity wave modes in RCE} 
As seen in Fig. \ref{fig: RCE}a), the RCE state of the STBL has a baseline vertical stratification profile. This, along with the choice of domain size, can excite background gravity wave modes independent of the applied forcing. A simple way to detect these modes is to compute the two-point autocorrelation of the velocity field components
\begin{equation}\label{eq: Rcorr}
    R_{ii}(r) = \ov{u'_i(x_i)u'_i(x_i+r)}
\end{equation} where the overline denotes averaging in the homogeneous dimensions and $r$ is a separation distance in a particular homogeneous direction. If $R_{ii}$ decays to 0 by the half-domain length, then the turbulent statistics are sufficiently decorrelated and thus the box size is adequate \citep{km87}. 
\begin{figure}
    \begin{center}
    \includegraphics[width=0.9\textwidth]{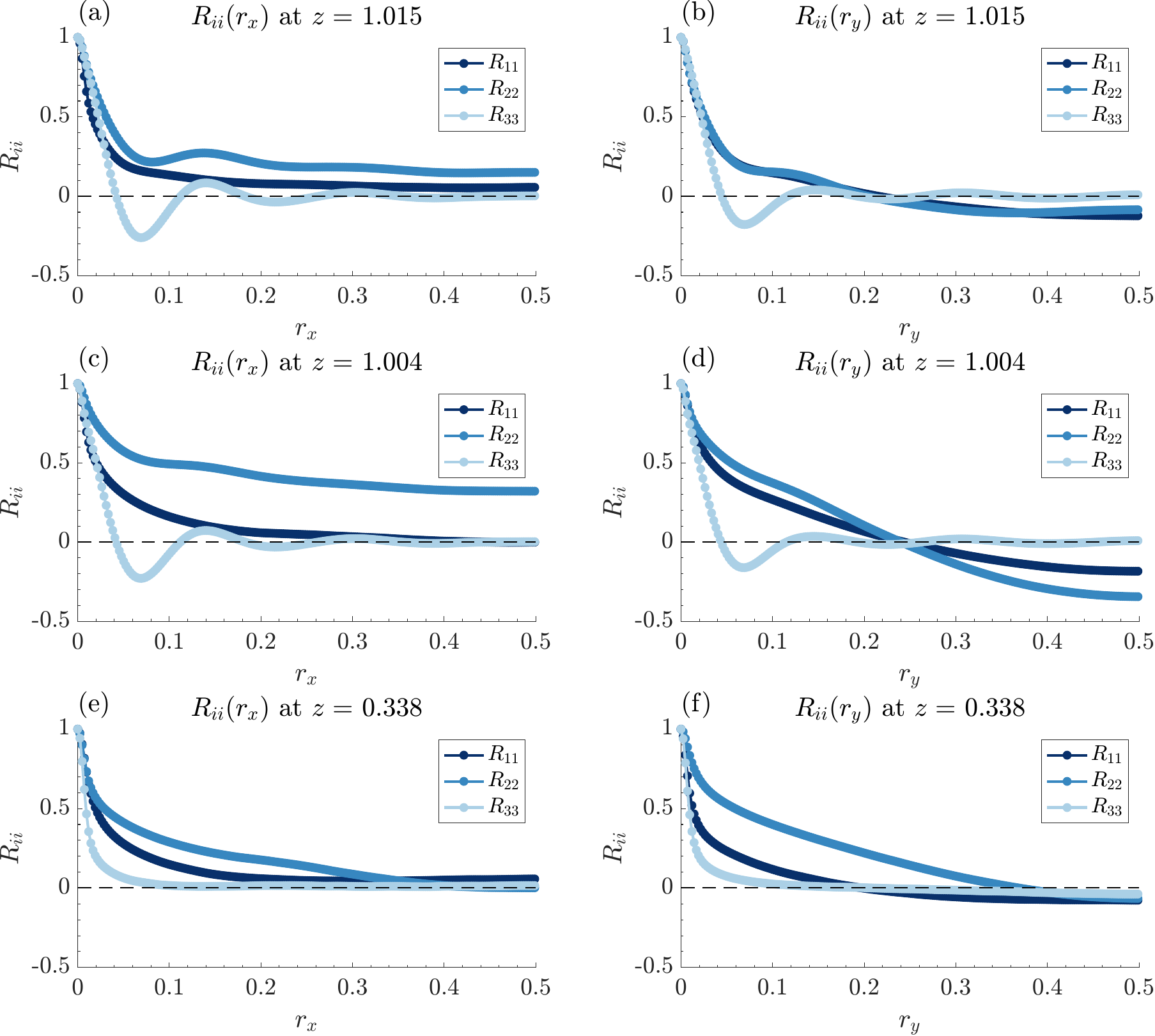}
    \caption{Two-point autocorrelations normalized by the variance as a function of $x$ and $y$ within the strongly stratified cloud layer (a-d) and outside of it (e-f) at three $z$ locations. \label{fig: R}}
    \end{center}
\end{figure} Notice that Fig. \ref{fig: R} largely illustrates the lack of decay of the horizontal autocorrelations for $z>1$ (i.e., above the inversion layer). This behavior is not remedied by grid refinement and/or increasing $L_x$ and $L_y$, suggesting the existence of background waves, as any strong modal signature in the flow (supported by the domain) will always be correlated with itself at some point downstream. 

\begin{figure}
    \begin{center}
    \includegraphics[width=0.9\textwidth]{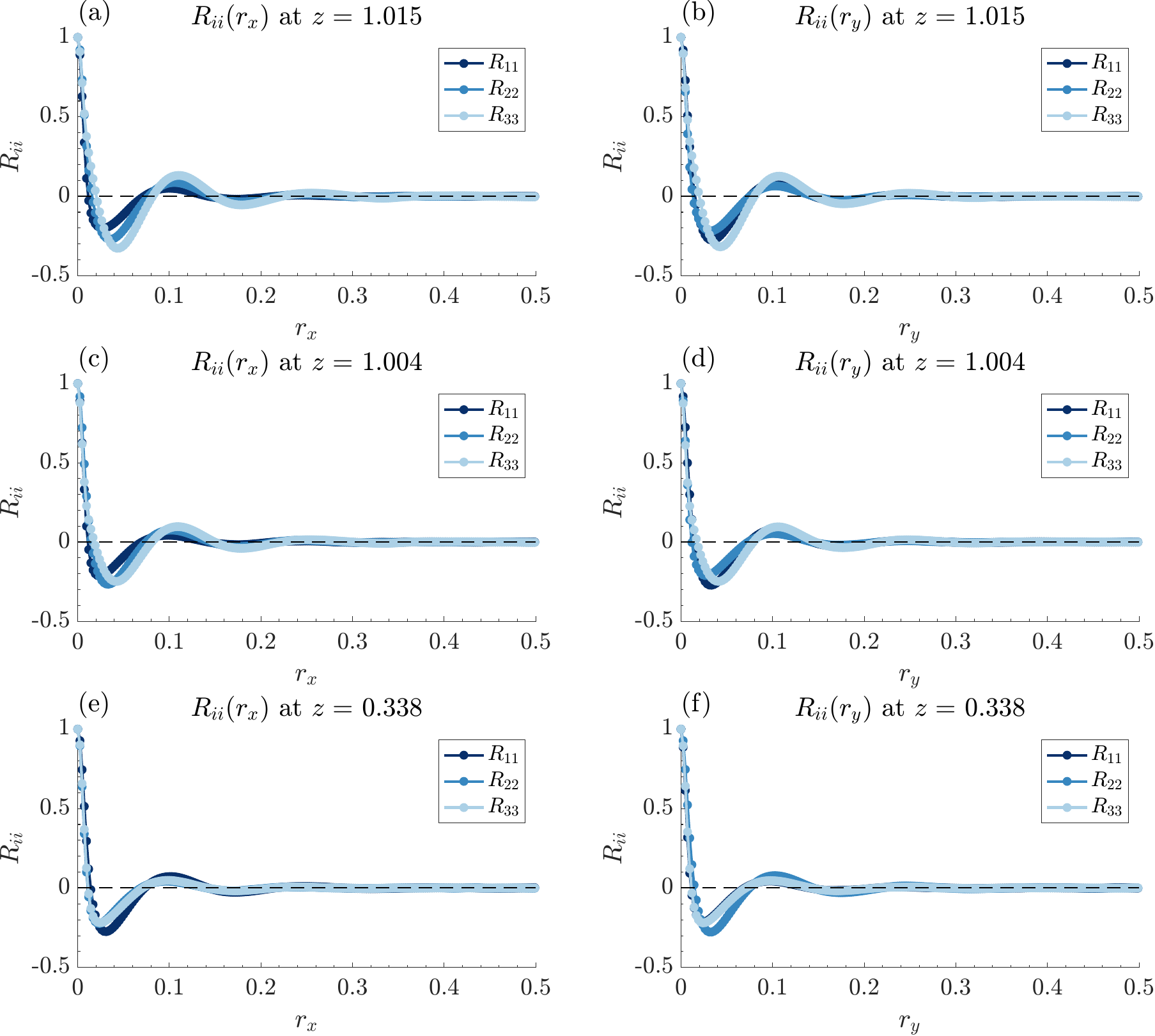}
    \caption{Same as Fig. \ref{fig: R} but with a high-pass filter applied to the fluctuating velocity fields. \label{fig: Rhpf}}
    \end{center}
\end{figure}

In order to distinguish between (large-scale) wave modes and the underlying turbulent state, we compute the energy spectra, which is the Fourier transform of Eq. \ref{eq: Rcorr}. 
\begin{figure}
    \begin{center}
    \includegraphics[width=0.9\textwidth]{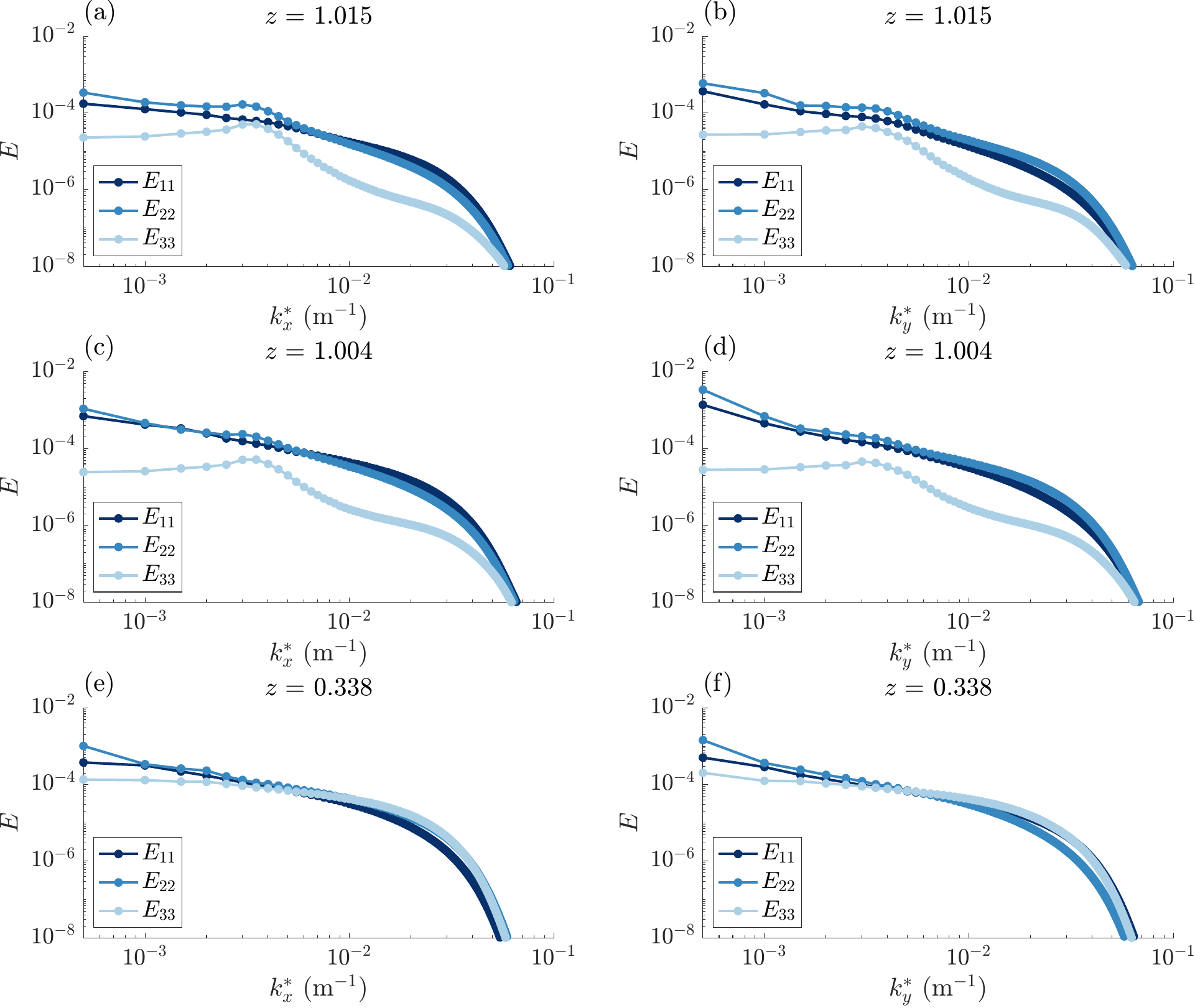}
    \caption{Energy spectra as a function of horizontal wavenumber at the same three vertical locations as Fig \ref{fig: R}. \label{fig: E}}
    \end{center}
\end{figure} Here in Fig. \ref{fig: E}, as we enter the stably stratified portion of the STBL ($z>1$), there is a peak in $E_{22}$ and $E_{33}$ at $k^*_x = k^*_y = 0.0035$~m$^{-1}$. This is suggestive of a critical wavenumber ($k^*_c = 0.0035$~m$^{-1}$) that demarcates these wave modes from the general turbulent behavior of the STBL. Further, this $k^*_c$ seems to be robust across vertical locations in the region of stable stratification. Applying a high-pass filter at this critical wavenumber on the velocity fields leads to the convergence of the autocorrelation and thus illustrates the appropriateness of this choice of domain size (Fig. \ref{fig: Rhpf}). Note that the presence of these background gravity waves did not impact stationarity as evinced by Fig. \ref{fig: RCE}b) as their energy is far lower (Fig. \ref{fig: E}) than what is required to cause breakup. 

\appendix[C]
\appendixtitle{Turbulent kinetic energy budget closure}
Derivation of the TKE equations begins via a Reynolds decomposition of the flow field: $u_i^* = \ov{u}_i^* + u_i'^*$. For simplicity, we will drop the $^*$, but we are still referring to dimensional quantities. In the momentum equation, it is convenient to label the individual terms as
\begin{equation}\label{eq: momredef}
    \f{\pd u_i}{\pd t} =  -\ub[1pt]{u_j\f{\pd u_i}{\pd x_j}}_\text{\clap{$A_i$~}} \ub[1pt]{-c_p\theta_\rho \f{\pd \pi'}{\pd x_i}}_\text{\clap{$P_i$~}} - \ub[1pt]{2\epsilon_{ijk}\Omega_j(u_k-u^{G}_k)}_\text{\clap{$C_i$~}} + \ub[1pt]{\f{g}{\theta_{\rho 0}}\theta'_\rho}_\text{\clap{$B$~}}\delta_{i3} + \mathcal{F}\delta_{i3} + N_i + W_i.
\end{equation} The mean equation is then 

\begin{equation}\label{eq: meanmom}
    \f{\pd \ov{u}_i}{\pd t} = -\ov{A}_i + \ov{P}_i - \ov{C}_i + \ov{B}\delta_{i3} + \ov{\mathcal{F}}_i\delta_{i3} + \ov{T}_i + \ov{N}_i + \ov{W}_i. 
\end{equation} Subtracting Eq. \ref{eq: meanmom} from Eq. \ref{eq: momredef} yields the turbulent momentum equation
\begin{equation}\label{eq: turbmom}
    \f{\pd u_i'}{\pd t} = -A_i' + P_i' - C_i' + (B'+\mathcal{F'})\delta_{i3} + T_i' + N_i' + W_i'. 
\end{equation} Defining the TKE as $k = \f{1}{2}\ov{u_i'u_i'}$, we can multiply Eq. \ref{eq: turbmom} by $u_i'$ and average over the periodic dimensions (and time if applicable). Eq. \ref{eq: tkeapp} is expressed in this form to minimize any algebraic manipulations that may not hold discretely in order to ensure $\pd k / \pd t$ is equal to the sum of right-hand-side terms (what we will call the TKE tendency).  
\begin{equation}\label{eq: tkeapp}
    \f{\pd k}{\pd t} = -\ov{u'_iA'_i} + \ov{u'_iP'_i}+ (\ov{u'_iB'}+\ov{u'_i\mathcal{F}'})\delta_{i3} + \ov{u'_iT'_i} + \ov{u'_iN'_i} + \ov{u'_iW'_i}
\end{equation} Note that the Coriolis contribution to TKE is identically zero. The various terms on the right-hand side of Eq. \ref{eq: tkeapp} are computed via the domain diagnostic branch of CM1 (see the ``resolved tke budget'' variables). Of particular note is that the various budget terms ($A_i$, $P_i$, etc.) are taken directly from the dynamic portion of the solve routines such that the user-selected discretization scheme is preserved. However, the budget terms are saved upon the last step of the RK3 routine, meaning that the terms themselves are determined via quantities evaluated at the second substep of the time stepping routine. This necessitates an average in time between the ``$n$'' and ``$n+1$'' steps to compute $u_i'$. Closure is not achieved otherwise if no interpolation in time is performed.
\begin{figure}
    \begin{center}
    \includegraphics[width=0.6\textwidth]{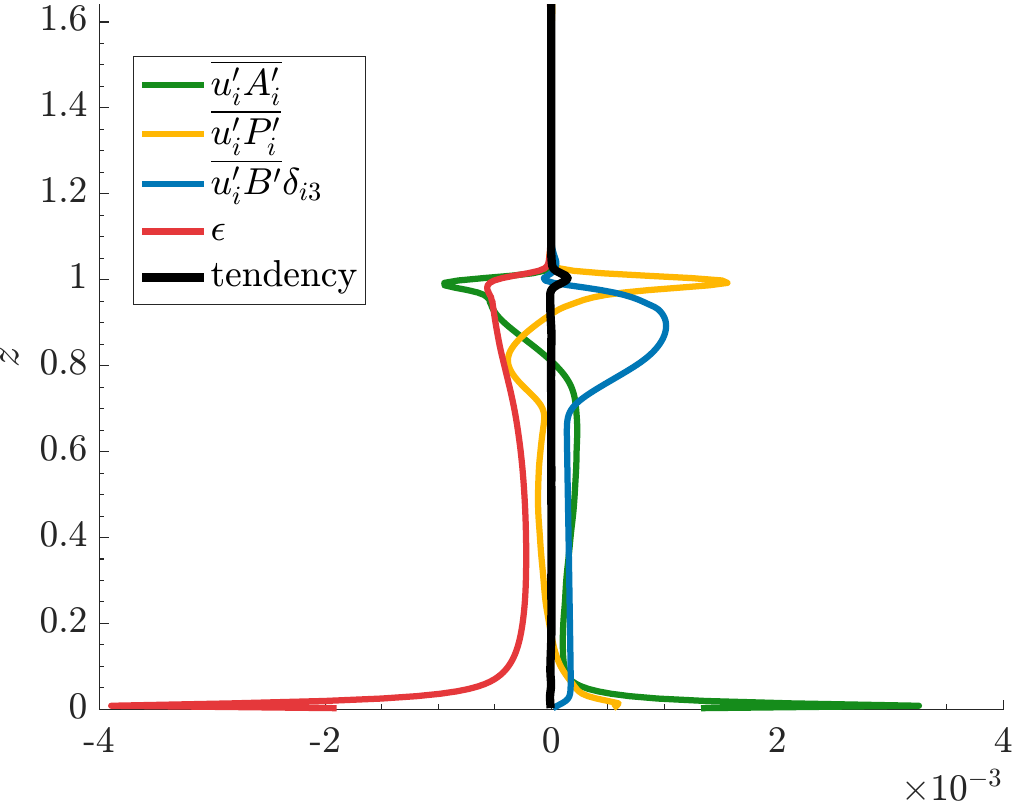}
    \caption{TKE budget for the RCE state as a function of $z$. Terms were averaged in the horizontal directions as well as in time. The black ``tendency'' line denotes the sum of all budget terms. \label{fig: RCEtke}}
    \end{center}
\end{figure} Fig. \ref{fig: RCEtke} illustrates the budget terms as a function of height for the RCE simulation. In the stationary state, we expect the change in TKE to be zero, which should be reflected by the sum of the right-hand side of Eq. \ref{eq: tkeapp}. The budget terms were calculated at a frequency of $\Delta t^* = 0.5$~s and averaged over a time period of $1.5$~hr. We see that the closure level is excellent as the tendency term is $O(10^{-5}$), which is two orders of magnitude smaller than the various terms in the budget. There is a slight deviation from zero near the inversion layer, but that is attributable to the error caused by interpolation for conformal operations on a staggered grid. The error reduces over longer time averaging and calculating budget terms at shorter output frequencies. Note that the dissipation $\epsilon$ is the sum of that due to the SGS model and sponge near the boundaries. 

\begin{figure}
    \begin{center}
    \includegraphics[width=0.6\textwidth]{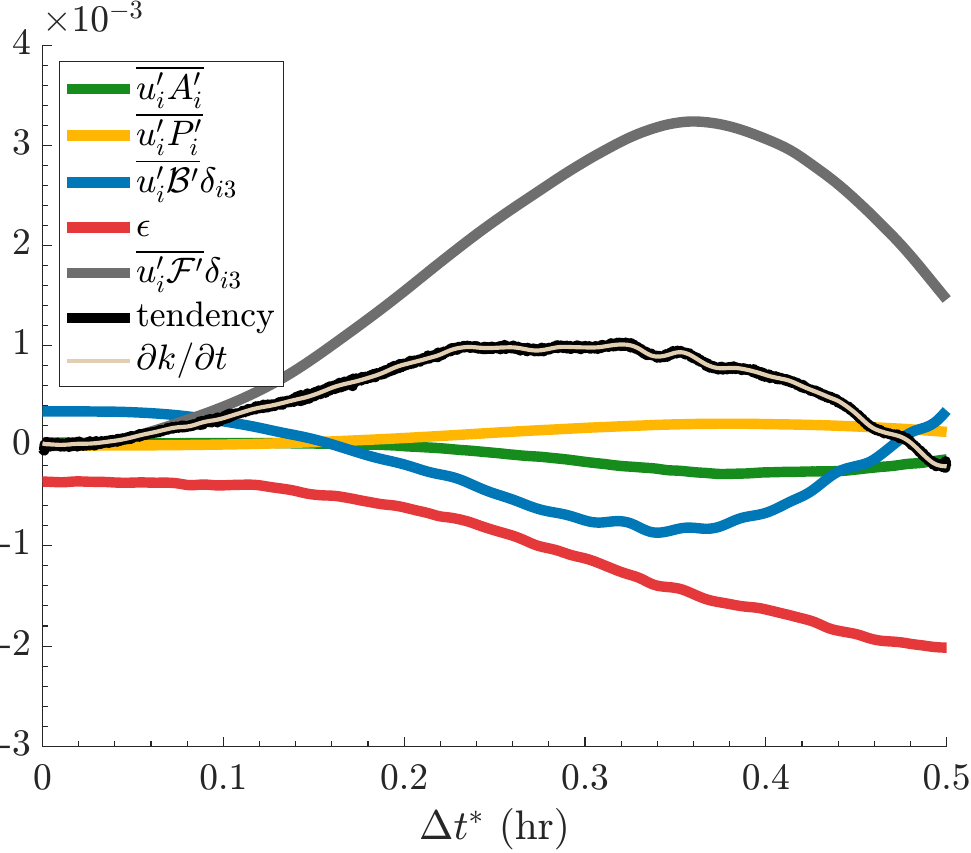}
    \caption{TKE budget for case A3. Here, a volume average was performed, and the closure is evaluated by comparing the black tendency and thin blue $\pd k / \pd t$ lines, which agree quite well. \label{fig: A3tke}}
    \end{center}
\end{figure} In the case of a non-stationary STBL, we can demonstrate a similar level of closure for one of the forced experiments. Fig. \ref{fig: A3tke} shows the agreement between the sum of budget terms (which now includes the contribution of the forcing and verifies its implementation) and the growth of TKE with time.  


\bibliographystyle{ametsocV6}
\bibliography{references}

\end{document}